\documentclass[AMA,STIX1COL]{WileyNJD-v2}

\usepackage{bm}
\newcommand\ie{\textit{i.e. }}
\definecolor{cinnamon}{rgb}{0.82, 0.41, 0.12}

\articletype{RESEARCH ARTICLE}%

\received{XXX 2018}
\revised{XXX 2018}
\accepted{XXX 2018}

\begin{document}

\title{Computational modeling of multiphase viscoelastic and elastoviscoplastic flows}

\author[1]{Daulet Izbassarov*}
\author[1]{Marco E. Rosti}
\author[1]{M. Niazi Ardekani}
\author[2]{Mohammad Sarabian}
\author[2]{Sarah Hormozi}
\author[1]{Luca Brandt}
\author[1]{Outi Tammisola}%
\authormark{IZBASSAROV \textsc{et al}}
\address[1]{Linn$\acute{\rm e}$ Flow Centre and SeRC (Swedish e-Science Research Centre), KTH Mechanics, SE 100 44 Stockholm, Sweden}
\address[2]{Department of Mechanical Engineering, Ohio University, Athens, OH, 45701-2979, USA}
\corres{*Izbassarov Daulet, Linn$\acute{\rm e}$ Flow Centre and SeRC (Swedish e-Science Research Centre), KTH Mechanics, SE 100 44 Stockholm, Sweden. 
\email{daulet@mech.kth.se}}

\abstract[Summary]{In this paper, a three-dimensional numerical solver is developed for suspensions of rigid and soft particles and droplets in viscoelastic and elastoviscoplastic (EVP) fluids. 
The presented algorithm is designed to allow for the first time three-dimensional simulations of inertial and turbulent EVP fluids with a large number particles and droplets. 
This is achieved by combining fast and highly scalable methods such as an FFT-based pressure solver, with the evolution equation for non-Newtonian (including elastoviscoplastic) stresses. 
In this flexible computational framework, the fluid can be modelled by either Oldroyd-B, neo-Hookean, FENE-P, and Saramito EVP models, and the additional equations for the non-Newtonian stresses 
are fully coupled with the flow. The rigid particles are discretized on a moving Lagrangian grid while the flow equations are solved on a fixed Eulerian grid. The solid particles are represented 
by an Immersed Boundary method (IBM) with a computationally efficient direct forcing method allowing simulations of a large numbers of particles. The immersed boundary force is computed at the 
particle surface and then included in the momentum equations as a body force. The droplets and soft particles on the other hand are simulated in a fully Eulerian framework, the former with a 
level-set method to capture the moving interface and the latter with an indicator function. The solver is first validated for various benchmark single-phase and two-phase elastoviscoplastic flow 
problems through comparison with data from the literature. Finally, we present new results on the dynamics of a buoyancy-driven drop in an elastoviscoplastic fluid.}


\maketitle


\section{Introduction} \label{intro}
Elastoviscoplastic (EVP) fluids can be found in geophysical applications, such as mudslides and the tectonic dynamic of the Earth. EVP fluids are also found
in industrial applications such as mining operations, the conversion of biomass into fuel, and the petroleum industry, to name a few. Biological and smart materials can be elastoviscoplastic, making the EVP fluid flows relevant for problems in physiology,
biolocomotion, tissue engineering, and beyond. In most of these applications, we are dealing with multiphase flows \cite{maleki2017submerged,gholami2017time,firouzniainteraction,hormozi2014visco,
hormozi2012nonlinear,liu2016two,LiuAxisymmetric}. Therefore, there is a compelling need to study multiphase flows of EVP fluids and predict their flow dynamics in various situations, including three-dimensional and inertial flows. 
Elastoviscoplastic materials exhibit simultaneously elastic, viscous and plastic properties. At low strains the material exhibits elastic 
deformation, whereas at sufficiently high strains the material experiences irreversible deformation and starts to flow. 
Even conventional  yield-stress  test fluids (such as Carbopol solutions and liquid foams) are shown to exhibit simultaneously elastic, viscous and yield stress behavior. 
Hence, in order to accurately predict the behavior 
of such materials, it is essential to model them as an EVP fluid, rather than an ideal yield-stress fluid (e.g., using the Bingham or the Herschel-Bulkley models).  

There are different types of models that have been proposed for EVP fluids. For instance, Saramito \cite{saramito2007new} proposed a tensorial constitutive law under the Eulerian framework, 
which is based on the combination of the Bingham viscoplastic \cite{bingham1922fluidity,herschel1926measurement} and the Oldroyd viscoelastic models \cite{oldroyd1950formulation} in a way 
which satisfies the second law of thermodynamics. This model predicts a Kelvin-Voigt viscoelastic solid (an ideal Hookean solid) response before yielding, when the von Mises criterion 
is not satisfied. Once the strain energy exceeds a threshold value that is specified by the von Mises criterion, the material yields, and the stress field is given by the non-linear viscoelastic constitutive law. 
This model was later improved by the same author \cite{saramito2009new} to account for the shear-thinning behavior of the shear viscosity, and also for the smoothness of the plasticity criterion. 
Moreover, this model is capable of predicting the first normal stress difference along with the yield stress behavior in simple shear flows as a result of combining viscoelasticity and viscoplasticity. 

The prediction of an ideal Hookean solid of Saramito's models \cite{saramito2007new,saramito2009new} for the EVP material before yielding causes the model to always predict a zero phase difference between the strain oscillation and the material shear stress, which in turn contributes to vanishing viscous harmonics. This results in an erroneous prediction of zero loss modulus $(G^{\prime \prime})$, 
which is in disagreement with the large amplitude oscillatory shear (LAOS) experiments for identifying and characterizing the properties of the EVP materials \cite{ewoldt2008new,ewoldt2010large}. 
It was shown by Dimitriou and co-workers \cite{dimitriou2013describing} that for a Carbopol gel (an EVP material), in the limits of small deformation amplitudes, the loss modulus  $(G^{\prime \prime})$ 
is always non-zero and indeed is smaller than the storage modulus $(G^{\prime})$ by an order of magnitude. The Isotropic Kinematic Hardening (IKH) idea was then suggested by Dimitriou and co-workers \cite{dimitriou2013describing} and Dimitriou and McKinley \cite{dimitriou2014comprehensive} 
to tackle this problem and to specify the parameters of the models correctly. Based on this concept, the material yield stress builds up and evolves in time together with the flow field, 
where the steady state yield stress is determined via the back stress modulus (a new material parameter) and the deformation of microstructure (a hidden internal dimensionless evolution variable). 
By this method, the energy is allowed to be dissipated, and thus, at small strain amplitudes, it predicts a non-vanishing loss modulus. Recently, a comprehensive IKH constitutive framework has 
been developed to model the thixotropic behavior presents in some practical  EVP materials such as  waxy crude oils \cite{geri2017thermokinematic}.

De Souza Mendes \cite{de2007dimensionless} proposed another constitutive equation for EVP fluids. The basic idea of this model is to modify the classical version of the Oldroyd-B equation, 
where the constant parameters, i.e.\ the relaxation time $(\lambda_{1})$, the retardation time $(\lambda_{2})$ and the viscosity $(\eta)$, are replaced with functions of the deformation rate. 
This model reduces to the classical Oldroyd-B equation in the limit of zero shear rate for the unyielded material. Benito and co-workers \cite{benito2008elasto} presented another minimal, 
fully tensorial and rheological constitutive equation for EVP fluids. This model predicts the material behaviour as a viscoelastic solid,  capable of deforming substantially before 
yielding, and predicts a viscoelastic fluid after yielding. Moreover, based on the second law of thermodynamics this model has a positive dissipation. 
Recently, Fraggedakis \etal \cite{Fraggedakis_JNNFM_2016} performed a systematic comparison of these recently proposed EVP fluid models. The models were tested in simple viscometric flows and against 
available experimental data.

A significant number of numerical studies have analysed purely viscoelastic and purely viscoplastic fluids, but a very limited number accounted for EVP fluids in which neither elastic nor plastic 
effects are negligible. The main reason is that numerical simulations of EVP fluid flows are not a straightforward task due to the inherent non-linearity of the governing equations. 
Nevertheless, numerical simulations can provide quantitative information which is extremely difficult to access by experiments in EVP fluids (for example, velocity fields and stress fields, separated 
into different contributions), and also detailed understanding of the physics of the interaction between particles and droplets in EVP fluids. 

Numerical simulations have already helped to reveal elastic effects in liquid foams and Carbopol. First, Dollet and co-workers \cite{dollet2007two} performed experimental measurements for the flow of liquid foam around a circular obstacle, where they observed an overshoot of the velocity (so-called negative wake) behind the obstacle. Then, Cheddadi \cite{cheddadi2011understanding} simulated the flow of an EVP fluid around a circular obstacle by employing Saramito's EVP model \cite{saramito2007new}. The numerical simulation using the EVP model captured the negative wake.  
A purely viscoplastic flow model (Bingham model) on the other hand always predicted fore-aft symmetry and the lack of a negative wake, in contrast with the aforementioned experimental observations. The numerical simulations could hence prove that the negative wake was an elastic effect.
Recently, the loss of the fore-aft symmetry and the formation of the negative wake around a single particle sedimenting in a Carbopol solution  was captured by  transient numerical calculations 
by Fraggedakis and co-workers \cite{Fraggedakis_SM_2016} by adopting the EVP tensorial constitutive law of Saramito \cite{saramito2007new}. This was in a quantitative agreement with the experimental 
observations by Holenberg and co-workers for the flow of Carbopol gel \cite{holenberg2012particle}. The elastic effects on viscoplastic fluid flows have also been addressed in numerical 
simulations of the EVP fluids through an axisymmetric expansion-contraction geometry \cite{nassar2011flow} by using the finite element method. It was observed that elasticity alters the shape and 
the position of the yield surface remarkably, and elasticity needs to be included to reach qualitative agreement with experimental observations for the flow of Carbopol aqueous solutions \cite{de2007flow}. 
Computations in the same geometry have also been performed by implementing the hybrid finite element-finite volume subcell scheme, and combining a regularization approach with the EVP model of 
Saramito \cite{saramito2007new}. Furthermore, Saramito model has been used to simulate the flow of liquid foam in a Taylor-Couette cell \cite{cheddadi2008numerical,cheddadi2012steady}. 
By adopting the EVP constitutive equation proposed by de Souza Mendes \cite{de2007dimensionless}, the flow pattern of EVP fluids in a cavity was investigated numerically, and it was demonstrated 
that the elasticity strongly affects the material yield surfaces \cite{martins2013elastic}. Recently, De Vita \etal \cite{fdv1} numerically investigated the elastoviscoplastic flow through porous media 
by adopting Saramito's model.

The motivation behind this work is to develop an efficient and scalable tool to deal with suspensions of particles and droplets in EVP fluids. In this work, we model an EVP fluid via the constitutive 
law proposed by Saramito \cite{saramito2007new}, which provided excellent results in previous numerical studies of \textit{e.g.} Carbopol, used in many experiments. 

Multiphase viscoelastic (EV) fluid flows have been studied much more than EVP flows, and indeed some of the results in literature will be used to validate our numerical implementation. To give a few examples of such studies, we list 2D and 3D direct numerical simulations of the dynamics of a rigid single particle \cite{huang1997direct,hwang2004direct,d2008rotation,
housiadas2011angular,villone2013particle}, two particles \cite{choi2010extended,yoon2012two,janssen2012collective, d2013dynamics}, multiple particles 
 \cite{santos2011alignment,hwang2011structure,de2013alignment,pasquino2014migration}, as well as 
droplets in viscoelastic two-phase flow systems in which one or both phases could be viscoelastic \cite{Izbassarov_JNNFM_2015,Izbassarov_POF_2016, Izbassarov_PRF_2016}, including the case of 
soft particles modeled as a neo-Hookean solid (\textit{i.e.}, a deformable particle is assumed to be a viscoelastic fluid with an infinite relaxation time) \cite{Villone_CF_2014,villone2014simulations2}. 

In the case of a pure visco-plastic (VP) suspending fluid, there is an abundance of computational studies of single and multiple particles \cite{beris1985creeping,blackery1997creeping,merkak2008dynamics, chaparian2017yield, 
chaparian2017cloaking}. Full 3D suspension flows for visco-plastic fluids are time consuming, and thus limited to a few benchmark calculations and lower mesh resolutions 
\cite{wachs2011peligriff, rahmani2014free}. 
However, 2D suspension flows are feasible \cite{yu2007fictitious}. 
The key computational challenge  is to resolve the structure of the unyielded regions, 
where the stress is below the yield stress, and to locate the yield surfaces that separate unyielded from yielded regions. Two basic methods are used: regularization and the Augmented Lagrangian 
(AL) approach \cite{frigaard2005usage}. Regularization tends to be faster, but may still require significant more resources than a Newtonian flow.  AL  approaches, although slower, properly 
resolve the stress fields. This is relevant for resolving important physical features of  suspensions of particles in visco-plastic fluids, e.g. the fact that buoyant particles can be held rigidly in suspension 
\cite{beris1985creeping, tokpavi2008very, putz2010creeping}, the limited influence of multiple particles on each other \cite{liu2003interactions}, and the finite arrest time, see Ref. \cite{maleki2015macro,Saramito_complex} for more details.
To overcome these limitations,  researchers have addressed yield stress suspensions from a 
continuum modeling closure perspective, deriving bulk suspension properties that agree with rheological experiments \cite{ovarlez2006local,coussot2009macroscopic, dagois2015rheology, 
ovarlez2015flows, hormozi2017dispersion}. 


The present manuscript is organized as follows. In the next section, the governing equations and the elastoviscoplastic constitutive models for multiphase elastoviscoplastic flows in complex geometries are 
briefly described. In Section \ref{sec: Num_method}, the numerical methodology is presented. In Section \ref{sec: Validation}, the numerical method is validated for various single-phase and two-phase elastoviscoplastic benchmark 
problems, and employed for buoyancy-driven elastoviscoplastic two-phase systems. In this work, we adopt two different IBM schemes to simulate EVP suspension flows which are  modifications and improvements 
of the original IBM scheme proposed by Peskin \cite{Peskin1972}. They are explained in section \ref{sec: Num_method} in more details. 
Finally some conclusions are drawn in Section \ref{sec: conclusion}. 

\section{Mathematical formulation}
\label{sec: gov eqns}
The dynamics of an incompressible flow of two immiscible fluids is governed by the Navier-Stokes equations, written in the non-dimensional form as:  
\begin{subequations}
 \begin{equation}
   \nabla \cdot {\bf u} = 0,
  \label{div free}
 \end{equation}
 \begin{equation}
   \rho \bigg( \frac{\partial \bf{u}}{\partial t} + {\bf u} \cdot \nabla {\bf u} \bigg) = -\nabla p 
+ \nabla \cdot \mu_s ( \nabla {\bf u} + \nabla {\bf u}^T ) + \nabla \cdot {\bm \tau} + \rho {\bf g} + {\bf f},
  \label{NS}
 \end{equation}
\label{mainNS}
\end{subequations}
where ${\bf u} = {\bf u} \left( {\bf x}, t \right)$ is the velocity field, $p = p \left( {\bf x}, t \right)$ is the pressure field, ${\bm \tau} = {\bm \tau} \left( {\bf x}, t \right)$ is an extra stress tensor (defined below) 
and $\bf{g}$ is a unit vector aligned with gravity or buoyancy. The term {\bf f} is a body force that is used to numerically impose the boundary conditions at the solid boundaries (particle-laden flow) and at the fluid-fluid interfaces (bubbly flow), as described in sections \ref{sec:bubbly} and \ref{sec:particle}. Finally, $\rho$ and $\mu_s$ are the density and the solvent viscosity of the fluid. 

In the present study, the viscoelastic and elastoviscoplastic effects in the flow are reproduced by the extra stress tensor $\bm{\tau}$. All the flow models (\textit{i.e.} the Neo-Hookean, viscoelastic Oldroyd-B, FENE-P and elastoviscoplastic Saramito model) can be expressed with a generic transport equation as 
\begin{equation}
\bigg(\frac{\partial {\bf B}}{\partial t}+{\bf u}\cdot \nabla {\bf B}- {\bf B}\cdot \nabla {\bf u}-\nabla {\bf u}^T\cdot {\bf B} \bigg)
 =\frac{a}{\lambda}{\bf I} - \frac{\mathcal{F}}{\lambda}{\bf B}. 
\label{SRM2}
\end{equation}
where $\lambda$ and $\mu_p$ are the relaxation time and polymeric viscosity, respectively. The definition of ${\bf B}$, $\mathcal{F}$ and $a$ used in Eq. \ref{SRM2} are specified in Table \ref{param} 
for the different models considered in the present study. In the Neo-Hookean material, $G$ is the shear elastic modulus; this model is analogous to considering the material as a viscoelastic fluid with an
infinite relaxation time $\lambda \rightarrow \infty$. In the Saramito model, $\bm{\tau}^d$ is the deviatoric stress tensor and its magnitude is defined as
\begin{equation}
|\bm\tau^d|= \sqrt{\frac{1}{2} \tau_{ij}^d\tau_{ij}^d}.
\end{equation}
In the FENE-P model, $L$ is the extensibility parameter defined as the ratio of the length of a fully extended polymer dumbbell to its equilibrium length.
From a numerical point of view, therefore, the challenges associated to the solution of equation (\ref{SRM2}) are similar, independently of the material model considered.

\begin{table}[h!]
\caption{Specification of the parameters ${\bf B}$, $\mathcal{F}$ and $a$ used in Eq. \ref{SRM2} for different models.}
\centering
\begin{tabular}{ l | c c c }
  \hline			
  Model & ${\bf B}$ & $\mathcal{F}$ & $a$ \\
  \hline			
  Neo-Hookean & $\bm{\tau}/G$ & 0 & 0 \\
  Oldroyd-B & $\bm{\tau} \lambda /\mu_p+\bm{I}$ & 1 & 1 \\
  Saramito &  $\bm{\tau} \lambda /\mu_p+\bm{I}$ & $\text{max}(0,1-\tau_y/|\bm{\tau}^d|)$ & $\mathcal{F}$ \\
  FENE-P & $(\bm{\tau} \lambda /\mu_p+ a \bm{I})/\mathcal{F}$ & $L^2/(L^2-trace({\bf B}))$ & $L^2/(L^2-3)$ \\
  \hline  
\end{tabular}
\label{param}
\end{table}

\section{Numerical method}
\label{sec: Num_method}
In this section, we outline the flow solver which has been previously developed for particle-laden flows \cite{Lambert_JFM_2013,Picano2015,Lashgari_IJMF_2016,Ardekani2016}, for bubbly flows \cite{Ge_AX_2017} and 
for viscoelastic flows \cite{Rosti_JFM_2017}. The grid is a staggered uniform Cartesian grid in which the velocity nodes are located at the cell faces, while the 
material properties, the pressure and the extra stresses are all located at the cell centers. The flow equations are solved 
using a projection method. The spatial derivatives are approximated 
using second-order central differences, except for the advection terms in Eqs. \eqref{SRM2}, \eqref{ls adv} and \eqref{reinit} where the 
fifth-order WENO or HOUC schemes are applied.

\subsection{Non-Newtonian fluid flow}
In a non-Newtonian flow, the transport equation for the extra stress tensor (Eq. \ref{SRM2}) presents specific challenges. Advection terms such as ${\bf u}\cdot \nabla {\bf B}$ need a special consideration due to the lack of diffusion terms in the equations. The most common approach is to 
introduce upwinding for the advection terms. However, that approach adds artificial dissipation that can cause the configuration tensor ${\bf B}$ to lose its positive definiteness, which eventually results 
in a numerical breakdown \cite{Joseph2013,Dupret1986}. Min \etal \cite{Min_JNNFM_2001} tested different spatial discretizations for a polymeric FENE-P fluid and showed that a third-order compact upwind scheme has a 
better performance. Dubief \etal \cite{Dubief2005} have also favored this scheme among the others. In both of these studies, a local artificial diffusion is added where the tensor ${\bf B}$ experiences a loss of positive definiteness $(det(B_{ij})<0)$. This discretization 
scheme works well, but is computationally expensive, because it requires to solve a set of linear equations for each component of the configuration tensor in each direction to calculate the derivatives. In this study we have substituted the compact upwind with an explicit fifth-order WENO scheme \cite{Liu_JCP_1994}, a considerably less expensive method that matches 
the performance of the compact scheme as the test case below illustrates; the method has been recently used successfully by Rosti and Brandt for an elastic material \cite{Rosti_JFM_2017}.

\begin{figure}[t]
\begin{center}
{\includegraphics[width=0.9\textwidth]{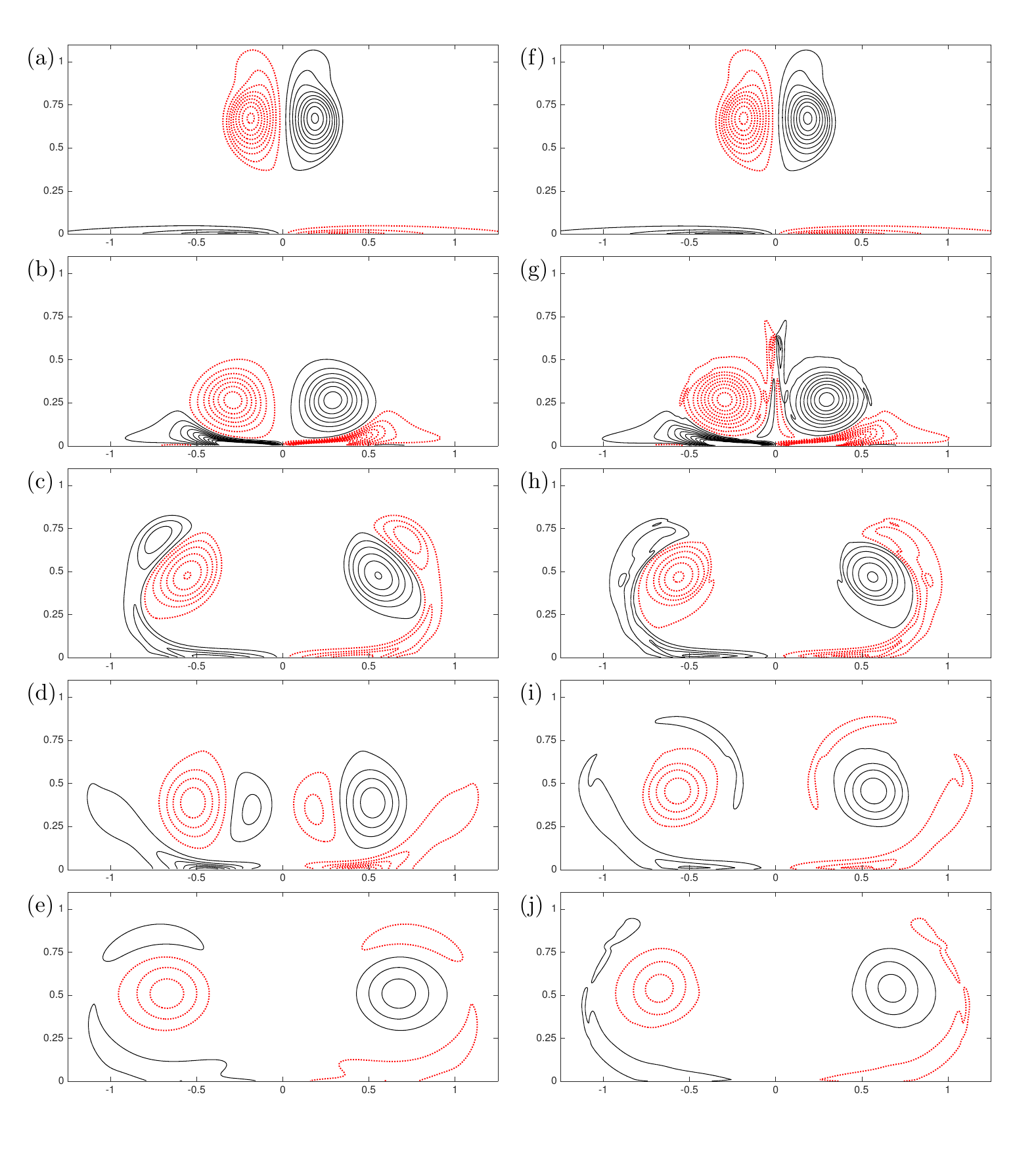}} 
\end{center}
\caption{Time sequence of vorticity contours for a two-dimensional vortex pair interacting with a wall at $t= 1$, $3$, $6$, $10$ and $15\delta^2/\Gamma_0$ for a Newtonian fluid ($a$-$e$) 
and a viscoelastic fluid ($f$-$j$). Contour levels are from -$15$ to $15$ with negative values indicated by red dashed lines.}
\label{fig:vo_pair}
\end{figure}

Next, we demonstrate the performance of our method in simulating a non-Newtonian fluid flow. A two-dimensional vortex pair interacting with a wall is simulated in a FENE-P fluid, similarly to Min \etal \cite{Min_JNNFM_2001}. In this flow, $Re$ and $Wi$ are defined as $Re = \rho \Gamma_0/(\mu_s + \mu_p) = 1800$ 
and $Wi = \lambda \Gamma_0/\delta^2 = 5$, where $ \Gamma_0$ is the initial circulation of the vortex and $\delta$ is the initial distance between the vortex pair center and the wall. 
The initial radius of each vortex is $0.145$ and the distance between the two centers is set to two radii. The solvent to total viscosity ratio 
$\beta=\mu_s/(\mu_s + \mu_p)$ is $0.9$ and the FENE-P extensibility parameter $L^2$ is $400$. 
Simulations are performed in a domain of size $2\delta \times 2\pi\delta$, with $64$ grid cells per $\delta$. Periodic boundary conditions are employed in the $x$-direction, and no-slip/no penetration boundary conditions are employed in the $y$ direction. 
A time sequence of the vorticity contours is shown in figure \ref{fig:vo_pair}, where the result for a Newtonian flow is also given as a reference. It can be observed that the secondary vortices 
are significantly attenuated in the polymeric flow.

\begin{figure}[htbp]
\begin{center}
{\includegraphics[width=0.9\textwidth]{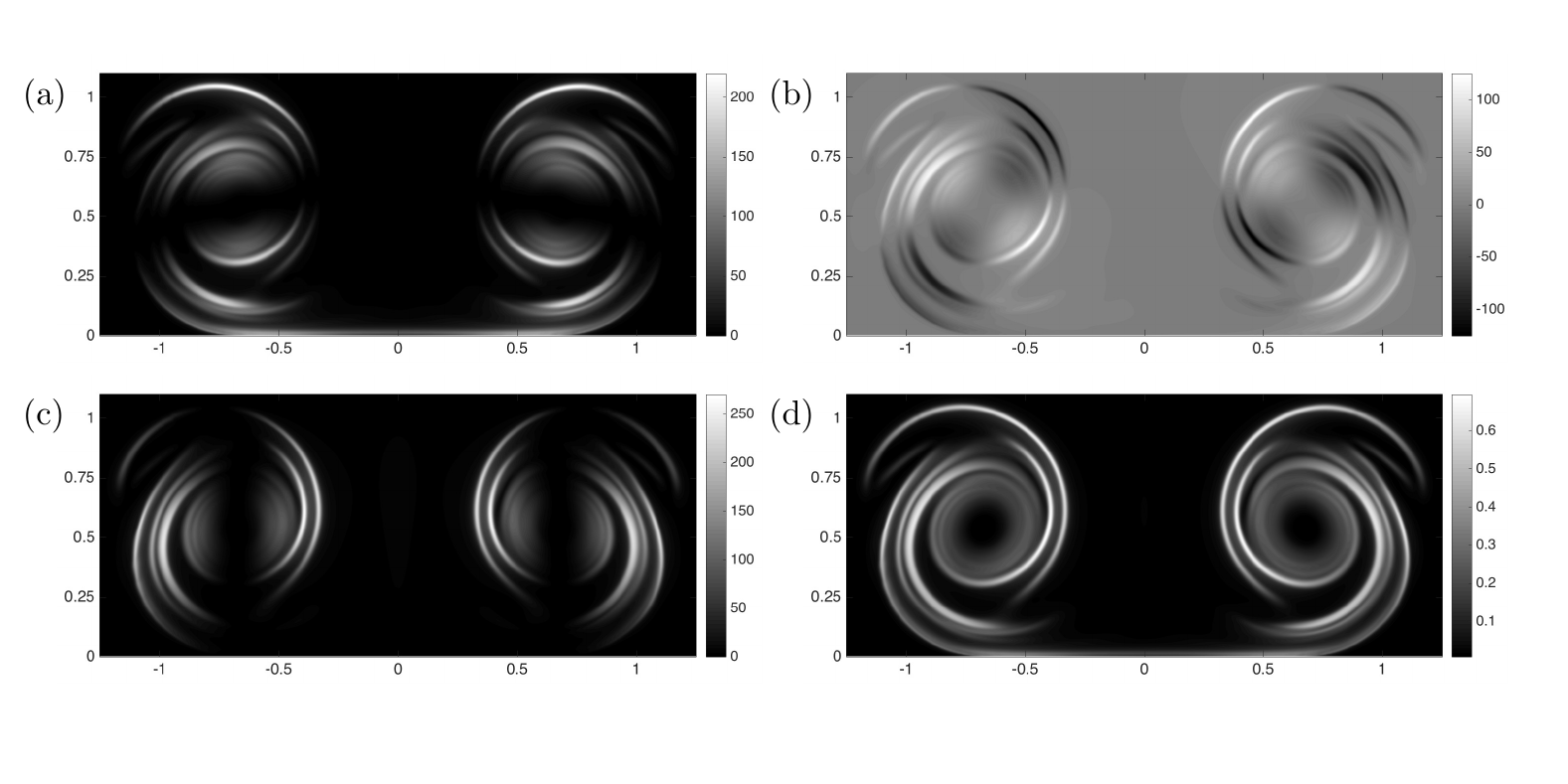}} 
\end{center}
\caption{Contours of $B_{xx}$ ($a$), $B_{xy}$ ($b$), $B_{yy}$ ($c$) and of  the trace of the tensor ${\bf B}$, normalized with $L^2$ ($d$) at $t = 15\delta^2/\Gamma_0$}
\label{fig:ContourPoly}
\end{figure}

A local artificial diffusion is added to the polymer equations \eqref{SRM2} in two instances: if the tensor ${\bf B}$ experiences a loss of positive definiteness $(det(B_{ij})<0)$, and if the trace of the tensor ${\bf B}$ 
reaches $95\%$ of its maximum (which is $L^2$). 
It is worth pointing out that in the case shown here, artificial diffusion was added in only a fraction of $0.1\%$ of the grid points. 
Contours of $B_{xx}$, $B_{xy}$, $B_{yy}$ and the trace of tensor ${\bf B}$, normalized with $L^2$ are given in figure \ref{fig:ContourPoly} at $t = 15\delta^2/\Gamma_0$. Adding the artificial diffusion to 
only a small fraction of grid points preserves the sharp spatial gradients of the tensor ${\bf B}$, as shown in this figure. The required amount of artificial diffusion needs to be tuned for each individual 
simulation as it changes with the relevant parameters of the polymeric flow; e.g. simulating the same test case here with $L^2=100$ removes any need for local artificial diffusion.

\subsection{Bubbly flow \label{sec:bubbly}}
\label{subsec: NS}
Fluid-fluid interfaces are captured by the interface-correction level-set method \cite{Ge_AX_2017},
and the surface tension force is described by the continuum surface force (CSF) model. The second-order Adams-Bashforth scheme (AB2) is used for the integration of governing equations of an EVP bubbly flow. Note that the AB2 scheme is used to facilitate the implementation of the fast pressure-correction
method developed by Dong and Shen \cite{Dong_JCP_2012} and Dodd and Ferrante \cite{Dodd_JCP_2014}.

\subsubsection{Level-set method}
In two-fluid systems, an interface between the phases can be resolved using a fully Eulerian method.  
The body force ${\bf f}$ due to surface tension, see Eqs. \eqref{mainNS}, is expressed as:
\begin{equation}
    {\bf f} = \sigma \kappa \delta(\phi) \bf n,
  \label{fLS}
\end{equation}
where $\delta$ is a regularized delta function and $\sigma$ is the surface tension.

In this paper, we have adopted a mass-conserving, interface-correction level-set method to capture an interface by a continuous level-set function. 
The level-set function $\phi({\bf x},t)$ approximates the signed distance from the interface. Hence, $\phi = 0$ denotes the interface, $\phi > 0$ denotes fluid 1 and $\phi < 0$  fluid 2. 
The interface is convected with the local velocity field, \ie 
\begin{equation}
  \frac{\partial \phi}{\partial t} + {\bf u} \cdot \nabla \phi = 0.
  \label{ls adv}
\end{equation}
To calculate the body force in Eq. \ref{fLS}, the unit normal vector, ${\bf n}$, and the local mean curvature, $\kappa$, can be simply computed as
\begin{equation}
    {\bf n} = \frac{\nabla \phi}{\vert \nabla \phi \vert },
  \label{normal}
\end{equation}
\begin{equation}
    \kappa = -\nabla \cdot {\bf n}.
  \label{curv}
\end{equation}

With time, if simply advected, the level set field will no longer equal a signed distance to the interface. It is essential that the signed distance 
property is preserved near the interface, because of the normal and curvature computation. We therefore redistance the level set field every 10-20 time 
steps by solving the Hamilton-Jacobi (reinitialization) equation:
\begin{equation}
{\frac{\partial \phi}{\partial {\mathcal T}}+S(\phi_0)(|\nabla \phi|-1)=0,}
\label{reinit}
\end{equation}
where $\phi_0$ is the level set field before redistancing, ${\mathcal T}$ is pseudo-time and $S(\phi_0)$ is the mollified sign function \cite{Ge_AX_2017}. 
One can observe that if a steady state is reached, then the zero level set contour (interface location) is unaltered, while the level set field has returned 
to a signed distance function. In practice, this equation is iterated only for a few steps towards steady state. 
The level set advection Eq. \eqref{ls adv} and reinitialization Eq. \eqref{reinit} are solved using a three-stage total-variation-diminishing (TVD) 
third-order Runge-Kutta scheme \cite{Shu_JCP_1988}.

The density, the solvent and the polymeric viscosities, and the relaxation time vary across the fluid interface and are expressed in a mixture form as
 \begin{equation}
 \begin{split}
  \rho &= \rho_1 H(\phi) + \rho_2 \big(1-H(\phi)\big), \quad
  \lambda = \lambda_{1} H(\phi) + \lambda_{2} \big(1-H(\phi)\big), \\ 
  \mu_s &= \mu_{s,1} H(\phi) + \mu_{s,2} \big(1-H(\phi)\big), \quad 
  \mu_p = \mu_{p,1} H(\phi) + \mu_{p,2} \big(1-H(\phi)\big),  
 \end{split}
 \end{equation}
 \label{mat prop}
where the subscripts $1$ and $2$ denote the properties of the bulk and suspended fluids, respectively, and $H(\phi)$ is the regularized Heaviside function defined such that it is zero inside the bubbles 
and unity outside.

A complete description of the level set methodology can be found in Sussman \etal \cite{Sussman_JCP_1994} and Ge \etal \cite{Ge_AX_2017}, 
and references therein.

\subsubsection{Time integration: Adams-Bashforth scheme\\}
To advance the solution from time level $n$ to $n+1$, we proceed as follows. First, we advance the level set function and update the density and viscosity fields accordingly. Second, we advance the extra stress tensor and the velocity field in time with the second-order Adams-Bashforth scheme as
\begin{equation}
    {\bf B}^{n+1} = {\bf B}^n+\Delta t \bigg(\frac{3}{2} {\bf RT_{ab}}^n-\frac{1}{2} {\bf RT_{ab}}^{n-1} \bigg),
  \label{ab2tau}
\end{equation}
\begin{equation}
    {\bf u}^* ={\bf u}^n +\Delta t \bigg(\frac{3}{2} {\bf RU_{ab}}^n-\frac{1}{2} {\bf RU_{ab}}^{n-1} \bigg),
  \label{ab2}
\end{equation}
where we have defined the right-hand sides of the Eqs. \eqref{SRM2} ${\bf RT_{ab}}^n$ and of eq. \eqref{NS} as ${\bf RU_{ab}}^n$, with
\begin{equation}
  {\bf RT_{ab}}^n = \bigg[\frac{1}{\lambda}\bigg( a {\bf I} -
  \mathcal{F}{\bf B}\bigg) - 
  \bigg({\bf u}\cdot \nabla {\bf B}-{\bf B}\cdot \nabla {\bf u}-
  \nabla {\bf u}^T\cdot {\bf B}\bigg) \bigg]^n, 
  \label{rt}
\end{equation}
\begin{equation}
\begin{aligned}
  {\bf RU_{ab}}^n = &-\nabla \cdot ({\bf u} {\bf u})^n + {\bf g} + \frac{1}{\rho^{n+1}}\bigg(\nabla \cdot \big[\mu_s^{n+1}(\nabla {\bf u}^n
+(\nabla {\bf u}^n)^T)\big] +\nabla \cdot \bm{\tau}^n 
  +\sigma \kappa^{n+1}\delta(\phi^{n+1}){\bf n}^{n+1} \bigg).
  \label{ru}
 \end{aligned}
\end{equation}
To enforce a divergence-free velocity field, Eq.\ \eqref{div free}, we proceed by solving the Poisson equation for the pressure \cite{Chorin_1968}, \ie
\begin{equation}
  \nabla \cdot \bigg(\frac{1}{\rho^{n+1}} \nabla p^{n+1} \bigg) = \frac{1}{\Delta t}\nabla \cdot {\bf u}^*,
  \label{var poisson}
\end{equation}
and finally, the velocity at the next time level is corrected as
\begin{equation}
    {\bf u}^{n+1} = {\bf u}^* -\frac{\Delta t}{\rho^{n+1}} \nabla p^{n+1}.
  \label{projection}
\end{equation}

In the droplet-laden flow, the pressure Poisson equation is solved in both phases, with unequal densities. Per default, the left hand side of the Poisson equation has variable coefficients. 
In order to utilise an efficient FFT-based pressure solver with constant coefficients \cite{Ge_AX_2017}, we use the following splitting of the pressure  term \cite{Dong_JCP_2012}:
\begin{equation}
{\frac{1}{\rho^{n+1}}\nabla p^{n+1} \to \frac{1}{\rho_0}\nabla p^{n+1}+\left(\frac{1}{\rho^{n+1}}-\frac{1}{\rho_0}\right)\nabla \left(2 p^{n}-p^{n-1}\right),}
\end{equation}
where $\rho_0$ is the density of the lower density phase (a constant). With this splitting, and after multiplying by $\rho_0$, the Poisson equation (Eq. \ref{var poisson}) becomes:
\begin{equation}
  \nabla \cdot \nabla p^{n+1} =  \nabla \cdot \left[\left(1-\frac{\rho_0}{\rho^{n+1}}\right)\nabla \left(2 p^{n}-p^{n-1}\right) \right]  + \frac{\rho_0}{\Delta t}\nabla \cdot {\bf u}^*.
 \label{poisson LS}
\end{equation}
and the velocity correction (Eq. \ref{projection}) transforms to:
\begin{equation}
    {\bf u}^{n+1} = {\bf u}^* -\Delta t\left[\frac{1}{\rho_0} \nabla p^{n+1}+\left(\frac{1}{\rho^k}-\frac{1}{\rho_0}\right) \nabla \left(2 p^{n}-p^{n-1}\right) 
\right].
 \label{projection LS}
\end{equation}

\subsection{Particle-laden flow \label{sec:particle}}
The governing equations of EVP particle-laden flow are integrated in time with a third order Runge-Kutta (RK3) scheme. The RK3 scheme is third order accurate, low storage, and improves the numerical 
stability of the code, allowing for larger time steps. Both rigid and deformable particles are considered here. The rigid particles are included using the immersed boundary method (IBM) that allows us to solve the Navier-Stokes equations on a Cartesian grid despite the presence of particles or complex wall geometries, and has become a popular tool in recent years.
The IBM consists of an extra force, added to the right-hand side of the momentum equations, see Eqs. \eqref{mainNS}, to mimic boundary conditions, creating virtual boundaries inside the numerical domain. This extra force acts in the vicinity of a solid surface to impose indirectly the 
no-slip/no-penetration (ns/np) boundary condition.

In the case of deformable particles, we use the method described in Rosti and Brandt \cite{Rosti_JFM_2017, rosti1, rosti2}. The solid is an incompressible viscous hyper-elastic material undergoing only the 
isochoric motion, where the hyper-elastic contribution is modeled as a neo-Hookean material, thus satisfying the incompressible Mooney-Rivlin law. To numerically solve the fluid-structure interaction problem, 
we adopt the so called one-continuum formulation \cite{tryggvason_sussman_hussaini_2007a}, where only one set of equations is solved over the whole domain. Thus, at each point of the domain the fluid and solid 
phases are distinguished by the local solid volume fraction $\phi_s$, which is equal to $0$ in the fluid, $1$ in the solid, and between $0$ and $1$ in the interface cells. The set of equations can be closed 
in a purely Eulerian manner by introducing a transport equation for the volume fraction $\phi_s$ (the equation is formally the same used in the level-set method, i.e., equation \eqref{ls adv}). 
The instantaneous local value of the elastic force is found by solving equation \eqref{SRM2}, which represents the upper convected derivative of the left Cauchy-Green deformation tensor. The right hand side of equation \eqref{SRM2} is identically 
zero for a hyperelastic material \cite{bonet_wood_1997a}.

\subsubsection{Time integration: Runge-Kutta scheme\\}
When using the third order Runge-Kutta scheme, the extra stress tensor and the unprojected field are computed by defining ${\bf RT_{rk3}}^k$ and ${\bf RU_{rk3}}^k$ as
\begin{equation}
\begin{aligned}
  {\bf RT_{rk3}}^k = & \zeta^{k} \bigg[\frac{1}{\lambda}\bigg(a{\bf I} -
  \mathcal{F}{\bf B}\bigg) - \bigg({\bf u}\cdot \nabla {\bf B}-{\bf B}\cdot \nabla {\bf u}- 
\nabla {\bf u}^T\cdot {\bf B}\bigg) \bigg]^{k-1} \\
                    + & \xi^{k} \bigg[\frac{1}{\lambda}\bigg(a{\bf I} -
  \mathcal{F}{\bf B}\bigg) - \bigg({\bf u}\cdot \nabla {\bf B}-{\bf B}\cdot \nabla {\bf u}- 
\nabla {\bf u}^T\cdot {\bf B}\bigg) \bigg]^{k-2}, 
  \label{RK-rt}
 \end{aligned}
\end{equation}
\begin{equation}
\label{eq:NS-num1}
\begin{split}
{\bf RU_{rk3}}^k = &- \zeta^{k} \nabla \cdot ({\bf u} {\bf u})^{k-1} - \xi^{k} \nabla \cdot ({\bf u} {\bf u})^{k-2} 
- 2 \frac{\alpha^{k}}{\rho} \nabla p^{k-1} +  \frac{\alpha^{k}}{\rho} \bigg(\nabla \cdot \big[\mu_s(\nabla {\bf u}+(\nabla {\bf u})^T)\big] + \nabla \cdot \bm{\tau} \bigg)^{k} \\ 
&+  \frac{\alpha^{k}}{\rho} \bigg(\nabla \cdot \big[\mu_s(\nabla {\bf u}+(\nabla {\bf u})^T)\big] + \nabla \cdot \bm{\tau} \bigg)^{k-1} ,
\end{split}
\end{equation}
which are the right-hand sides of the Eqs. \eqref{SRM2} and \eqref{NS}. Integrating in time yields 
\begin{equation}
    {\bf B}^{k} = {\bf B}^{k-1}+\Delta t {\bf RT_{rk3}}^k,
  \label{rk3tau}
\end{equation}
\begin{equation}
    {\bf u}^* ={\bf u}^{k-1} +\Delta t {\bf RU_{rk3}}^k.
  \label{rk3}
\end{equation}
In the previous equations, $\Delta t$ is the overall time step from $t^n$ to $t^{n+1}$, the superscript $^*$ is used for the predicted velocity, while the superscript $^k$ denotes 
the Runge-Kutta substep, with $k=0$ and $k=3$ corresponding to times $n$ and $n+1$. 

The pressure equation that enforces the solenoidal condition on the velocity field is solved via a Fast Poisson Solver
\begin{equation}
\label{eq:NS-num2}
\nabla \cdot \nabla \psi ^{k} = \frac{\rho}{2 \alpha^{k} \Delta t} \nabla \cdot {\bf u}^*,
\end{equation}
and, finally, the pressure and velocity are corrected according to
\begin{subequations}
\label{eq:NS-num3}
\begin{align}
p^{k} &= p^{k-1} + \psi ^{k}, \\
{\bf u}^{k} &= {\bf u}^{*} - 2 \alpha^{k} \frac{\Delta t}{\rho} \nabla \psi ^{k},
\end{align}
\end{subequations}
where $\psi$ is the projection variable, and $\alpha$, $\zeta$, and $\xi$ are the integration constants, whose values are
\begin{equation}
\label{eq:RK}
\begin{array}{ccc}
\alpha^1 = \frac{4}{15} & \alpha^2 = \frac{1}{15} & \alpha^3 = \frac{1}{6}\\ \\
\zeta^1 = \frac{8}{15} & \zeta^2 = \frac{5}{12} & \zeta^3 = \frac{3}{4}\\ \\
\xi^1 = 0 & \xi^2 = -\frac{17}{60} & \xi^3 = -\frac{5}{12}.
\end{array}
\end{equation} 

\subsubsection{Immersed boundary method}
The IBM force $\textbf{f}$ cannot be formulated by means of a universal 
equation and therefore IBMs differ in the way  $\textbf{f}$ is computed. The method applied in this solver was first developed by Peskin \cite{Peskin1972} and numerous modifications and improvements have been suggested since 
(for a review, see Ref. \cite{Mittal2005}). In this study we use two different IBM schemes: the volume penalization IBM \cite{Kajishima2001,Breugem2013} to generate obstacles and complex geometries, and the discrete forcing method for moving particles \cite{Breugem2012,Picano2015,Ardekani2016} to fully resolve particle suspensions in elastoviscoplastic flows.

\begin{figure}
\centering
\includegraphics[width=0.35\linewidth]{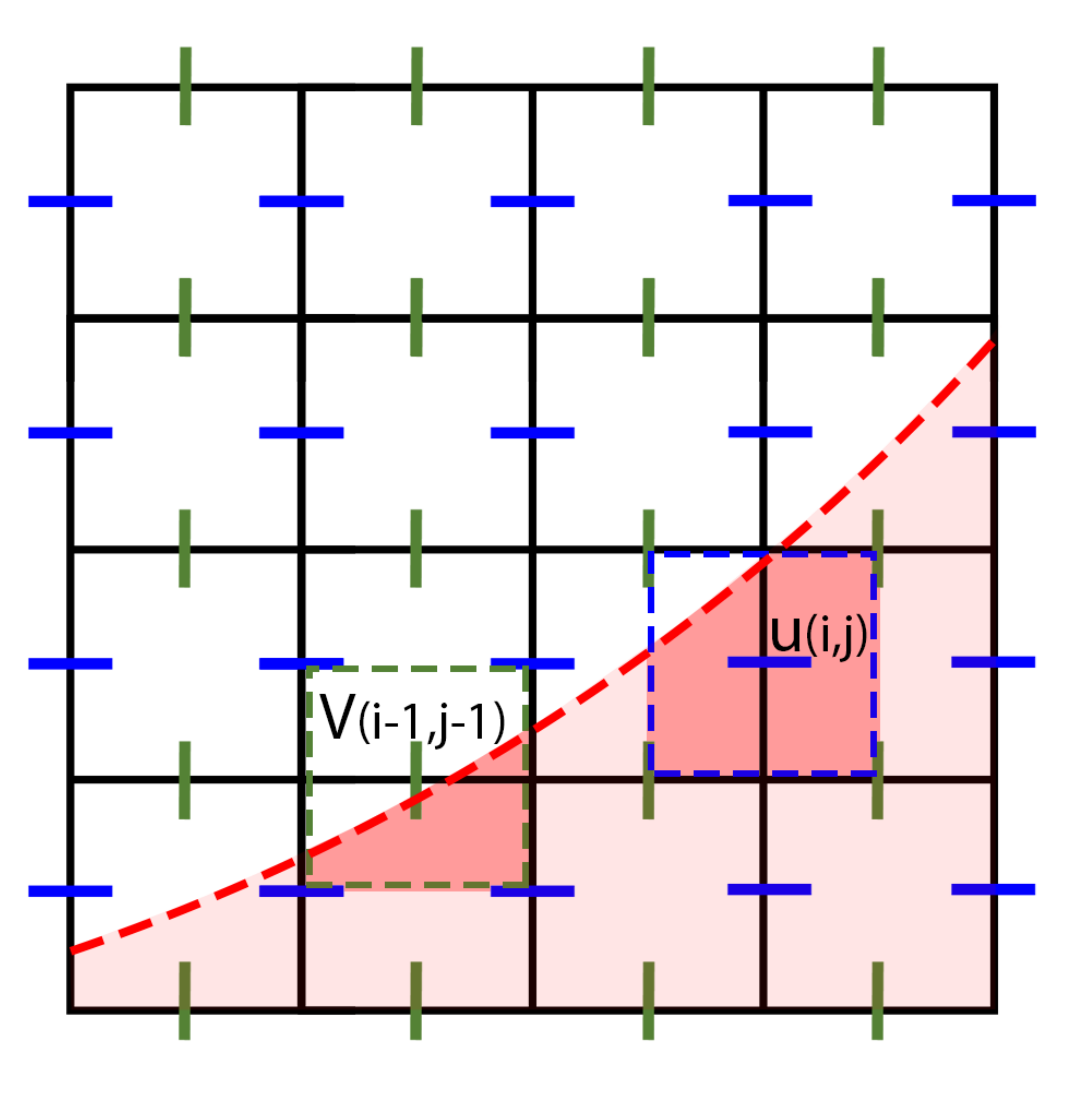}
\vspace{-20pt}
\caption{\label{fig:VolumeIBM}
Solid volume fractions (highlighted area) for grid cells around $u(i,j)$ and $v(i-1,j-1)$. Solid boundary is shown by red dashed line.}
\end{figure}

\paragraph*{Volume penalization IBM \\}
Kajishima \etal and Breugem \etal  \cite{Kajishima2001,Breugem2013} proposed the volume penalization IBM, where the IBM force $\textbf{f}$ is calculated from the first prediction velocity $\textbf{u}^{*}$ that is obtained by integrating Eq. \eqref{mainNS} in time without considering the IBM force and the pressure correction. The IBM force $\textbf{f}$ and the second prediction velocity $\textbf{u}^{**}$ are then calculated as follows:
\begin{subequations}
\label{eq:Vol}  
\begin{align}
 \textbf{f}_{ijk} \, =& \, \rho \, \alpha_{ijk} \frac{{\left(\textbf{u}_s - \textbf{u}^*\right)}_{ijk}}{\Delta t}  \,, \\[1em]
 \textbf{u}^{**}_{ijk} \, =& \, \textbf{u}^{*}_{ijk} \,+\, \Delta t \, \textbf{f}_{ijk} / \rho \, ,
 \end{align}
\end{subequations}
where $\alpha_{ijk}$ is the solid volume fraction in the grid cell with index $(i,j,k)$, varying between $0$ (entirely located in the fluid phase) and $1$ (entirely located in the solid area) and $\textbf{u}_s$ is the solid interface velocity within this grid cell. Figure \ref{fig:VolumeIBM} indicates the solid volume fractions (highlighted area) for grid cells around $u(i,j)$ and $v(i-1,j-1)$. Solid boundary in this figure is shown by red dashed line. For non-moving boundaries, $\textbf{u}_s$ is $0$ and Eq. \ref{eq:Vol} reduce to:
\begin{equation}
\label{eq:forceAdded}  
\textbf{u}^{**}_{ijk} = \left(1 - \alpha_{ijk} \right) \textbf{u}^{*}_{ijk}\,.
\end{equation}
The second prediction velocity $\textbf{u}^{**}$ is then used to update the velocities and the pressure following a classical pressure correction scheme \cite{Breugem2012}.

The volume penalization IBM is computationally very efficient, since the solid volume fractions around the velocity points can be calculated at the beginning of the simulation using an accurate method, or they can even be 
extracted directly from a physical sample by magnetic resonance imaging or X-ray computed tomography \cite{Breugem2013}. 

\begin{figure}
\centering
\includegraphics[width=0.35\linewidth]{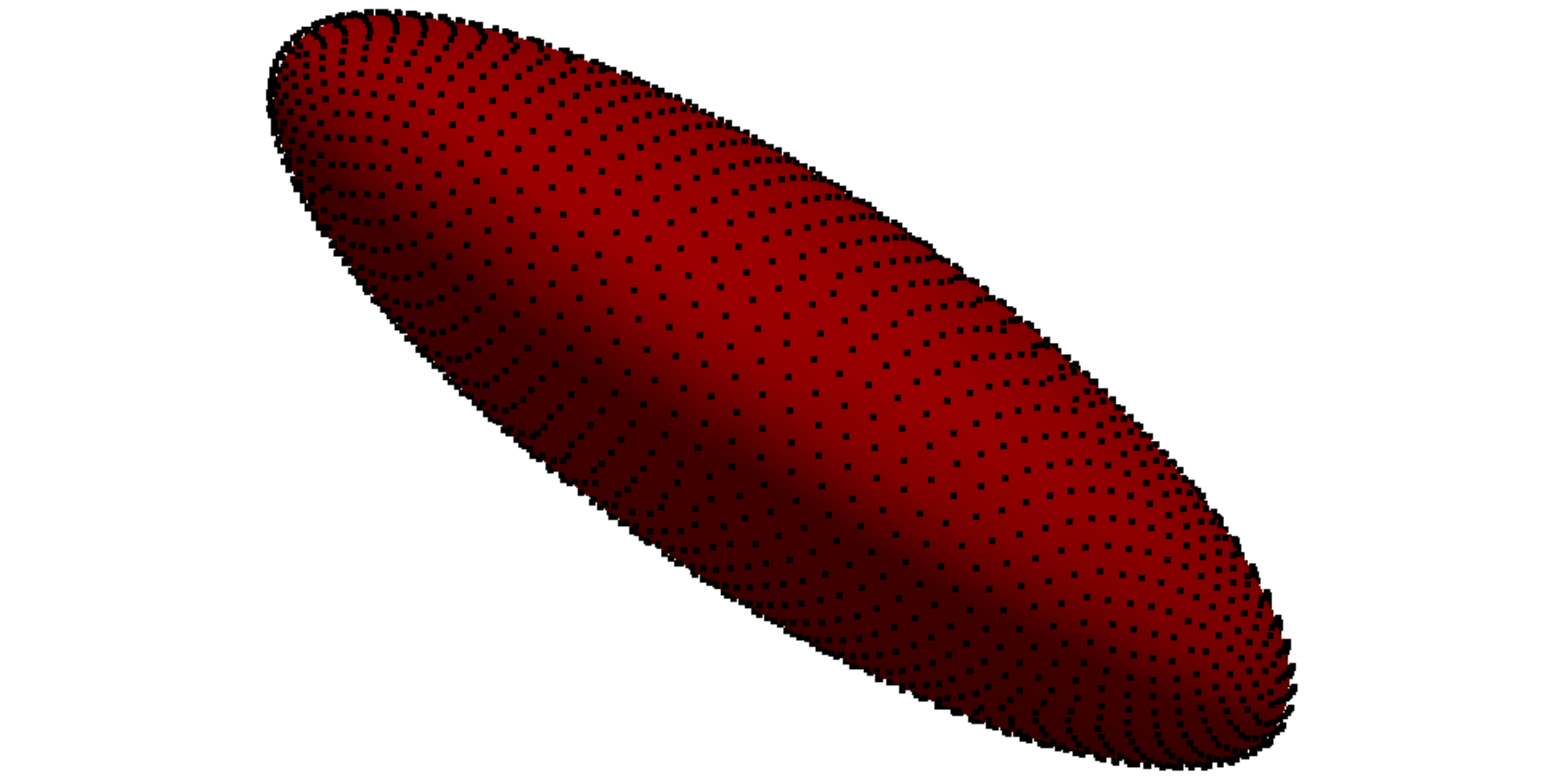}
\caption{\label{fig:Lag}
Uniform distribution of Lagrangian points over the surface of an spheroidal particle with an aspect ratio (polar over equatorial radius) $1/3$. }
\end{figure}

\paragraph*{Discrete forcing method for moving particles \\}
Following the IBM framework, we impose the no-slip/no penetration condition at the particle surfaces (Figure \ref{fig:Lag}) by adding an extra force $\textbf{f}$ on the right hand side of the fluid momentum equations \eqref{mainNS}.
Uhlmann \cite{Uhlmann2005} developed a computationally efficient IBM to fully resolve particle-laden flows. Breugem \cite{Breugem2012} introduced improvements to this method, making it second order accurate 
in space by applying a multi-direct forcing scheme \cite{Luo2007} to better approximate the no-slip/no-penetration (ns/np) boundary condition on the surface of the particles and by introducing a slight 
retraction of the grid points on the surface towards the interior. The numerical stability of this method for particle over fluid density ratio near unity was also improved by 
accounting the inertia of the fluid contained within the particles \cite{Kempe2012JCP}. Ardekani \etal \cite{Ardekani2016} extended the original method to simulate suspension of spheroidal particles  
with lubrication and contact models for the short-range particle-particle (particle-wall) interactions. 


In this study, we apply the same scheme to fully resolved simulations of particle suspensions in elastoviscoplastic flows.  We apply the IBM force on the predicted velocities $\textbf{u}^{*}$, 
which have been obtained as in the single-phase situation. The second prediction velocity $\textbf{u}^{**}$ is then obtained after the application of the IBM force, and $\textbf{u}^{**}$ 
substitutes $\textbf{u}^{*}$ in the pressure correction scheme given in the previous section. The formulation to calculate the second prediction velocity is given here:
\begin{subequations}
\label{eq:IBM}  
\begin{align}
&\textbf{U}^*_l = \sum\limits_{ijk} \textbf{u}^*_{ijk} \delta_d \left( \textbf{x}_{ijk} - \textbf{X}^{k-1}_l \right) \Delta x \Delta y \Delta z  \, , \\
&\textbf{F}^{k -1/2}_l = \rho_f \, \frac{\textbf{U} \left( \textbf{X}^{k-1}_l \right) - \textbf{U}^*_l}{\Delta t} \, , \\[1em] 
&\textbf{f}^{\, k -1/2}_{\, ijk} = \sum\limits_{l} \textbf{F}^{k-1/2}_l \delta_d \left( \textbf{x}_{ijk} - \textbf{X}^{k-1}_l \right) \Delta V_l \, , \\[0.6em]
&\textbf{u}^{**} = \textbf{u}^* + \Delta t \, \textbf{f}^{\, k -1/2} / \rho_f \, ,
\end{align}
\end{subequations}
where capital letters indicate the variable at a Lagrangian point with index $l$. In equation (\ref{eq:IBM}a), we interpolate the first prediction velocity $\textbf{u}^*$ from the Eulerian grid 
to the Lagrangian points on the surface of the particle, $\textbf{U}^*_l$, using the regularized Dirac delta function $\delta_d$ of Roma \etal \cite{Roma1999}. This approximated delta function 
essentially replaces the sharp interface with a thin porous shell of width $3\Delta x$; it preserves the total force and torque on the particle in the interpolation, provided that the Eulerian 
grid is uniform. The IBM force at each Lagrangian point, $\textbf{F}^{k-1/2}_l$, is proportional to the difference between the interpolated predicted velocity and the local velocity of the surface 
of the particle (for rigid particles, $\textbf{U}_p + \pmb{\omega}_p \times \textbf{r}$, calculated as shown in the paragraph below). In equation (\ref{eq:IBM}c), the IBM forces obtained at the 
Lagrangian points are interpolated back to the Eulerian grid by the same regularized Dirac delta function. In equation (\ref{eq:IBM}d), the IBM forces in the Eulerian grid ($\textbf{f}^{\, k -1/2}_{\, ijk}$) 
are added to the first prediction velocity to obtain the second prediction velocity $\textbf{u}^{**}$.

Given the smooth delta function and resolutions typically used, the Eulerian forces obtained from two neighboring Lagrangian points overlap. The \textit{multidirect forcing} scheme proposed by 
Luo \etal \cite{Luo2007} is therefore employed to iteratively determine the IBM forces such that the no-slip boundary conditions, $\textbf{U}^{**} \approx \textbf{U}$, are collectively imposed 
at the Lagrangian grid points. The new second prediction velocity $\textbf{u}^{**}$ is then obtained by solving the equations above iteratively (typically 3 iterations is enough) using the new 
$\textbf{u}^{**}$ as $\textbf{u}^{*}$ at the beginning of the next iteration with equation \ref{eq:IBM}b substituted by:
\begin{equation}
\label{eq:CoVe2} 
 \textbf{F}^{k -1/2}_l =  \textbf{F}^{k -1/2}_l + \rho_f \, \frac{\textbf{U} \left( \textbf{X}^{k-1}_l \right) - \textbf{U}^{**}_l}{\Delta t} \, .
\end{equation}

The second prediction velocity $\textbf{u}^{**}$ is then used to update the velocities and the pressure following the procedure described in the previous section.

Taking into account the inertia of the fictitious fluid phase inside the particle volumes, Breugem \cite{Breugem2012} showed that equations for particle motion can be rewritten as:
\begin{eqnarray}
\label{eq:NewtonEulerWim1}  
\rho_p V_p \frac{ \mathrm{d} \textbf{U}_{p}}{\mathrm{d} t} &\approx&  - \sum\limits_{l=1}^{N_L} \textbf{F}_l \Delta V_l + 
\rho_f  \frac{ \mathrm{d}}{\mathrm{d} t} \left( \int_{V_p} \textbf{u} \mathrm{d} V  \right)  + \left( \rho_p - \rho_f \right)V_p\textbf{g} + \textbf{F}_c , \, \\ [8pt] 
\frac{ \mathrm{d} \left( \textbf{I}_p \, \pmb{\omega}_{p} \right) }{\mathrm{d} t} &\approx& - \sum\limits_{l=1}^{N_L} \left(\textbf{r}_l \times \textbf{F}_l \right) \Delta V_l 
+ \rho_f  \frac{ \mathrm{d}}{\mathrm{d} t} \left( \int_{V_p} \left(\textbf{r} \times \textbf{u} \right) \mathrm{d} V  \right) + \textbf{T}_c  \, .
\label{eq:NewtonEulerWim2}  
\end{eqnarray}
where $\textbf{U}_p$ and  $\pmb{\omega}_{p}$ are the particle translational and the angular velocity, $\rho_p$, $V_p$ and $\textbf{I}_p$ are the mass density, volume and moment-of-inertia tensor 
of a particle, and $\textbf{r}$ is the position vector with respect to the center of the particle. The first terms on the right hand side of these equations are the summation of IBM forces and torques 
that act on each Lagrangian point. The second terms account for the translational and angular acceleration of the fluid trapped inside the particle shell. The force term $- \rho_f V_p\textbf{g}$ 
accounts for the hydrostatic pressure with $\textbf{g}$ the gravitational acceleration, and $\textbf{F}_c$ and $\textbf{T}_c$ are the force and torque resulting from particle-particle (particle-wall) 
collisions (see \cite{Ardekani2016} for more details). These equations are integrated in time using the Runge-Kutta scheme, as explained in the previous section.

\begin{figure}[htbp]
\begin{center}
{\includegraphics[width=0.9\textwidth]{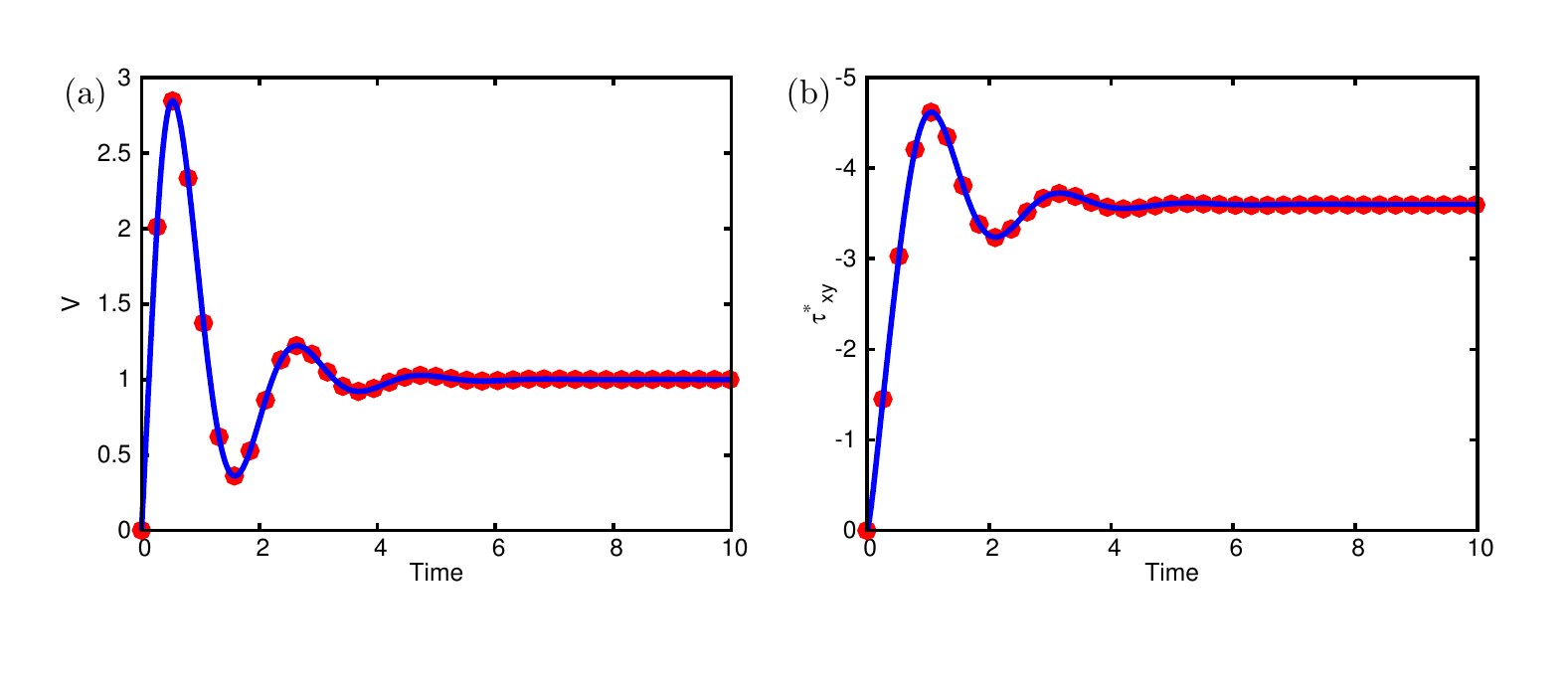}} 
\end{center}
\caption{Start-up Poiseuille flow for an Oldroyd-B fluid. (a) Time evolution of the centerline streamwise velocity component and (b) $\tau^*_{xy}$ stress at channel wall. 
The symbols represent our numerical results while the solid lines are the analytical solution derived by Waters and King \cite{Waters_RA_1970}. ($Re=0.125$, $Wi=0.125$ and $\beta=0.1$).}
\label{supfig1}
\end{figure}

\begin{figure}[htbp]
\begin{center}
{\includegraphics[width=0.9\textwidth]{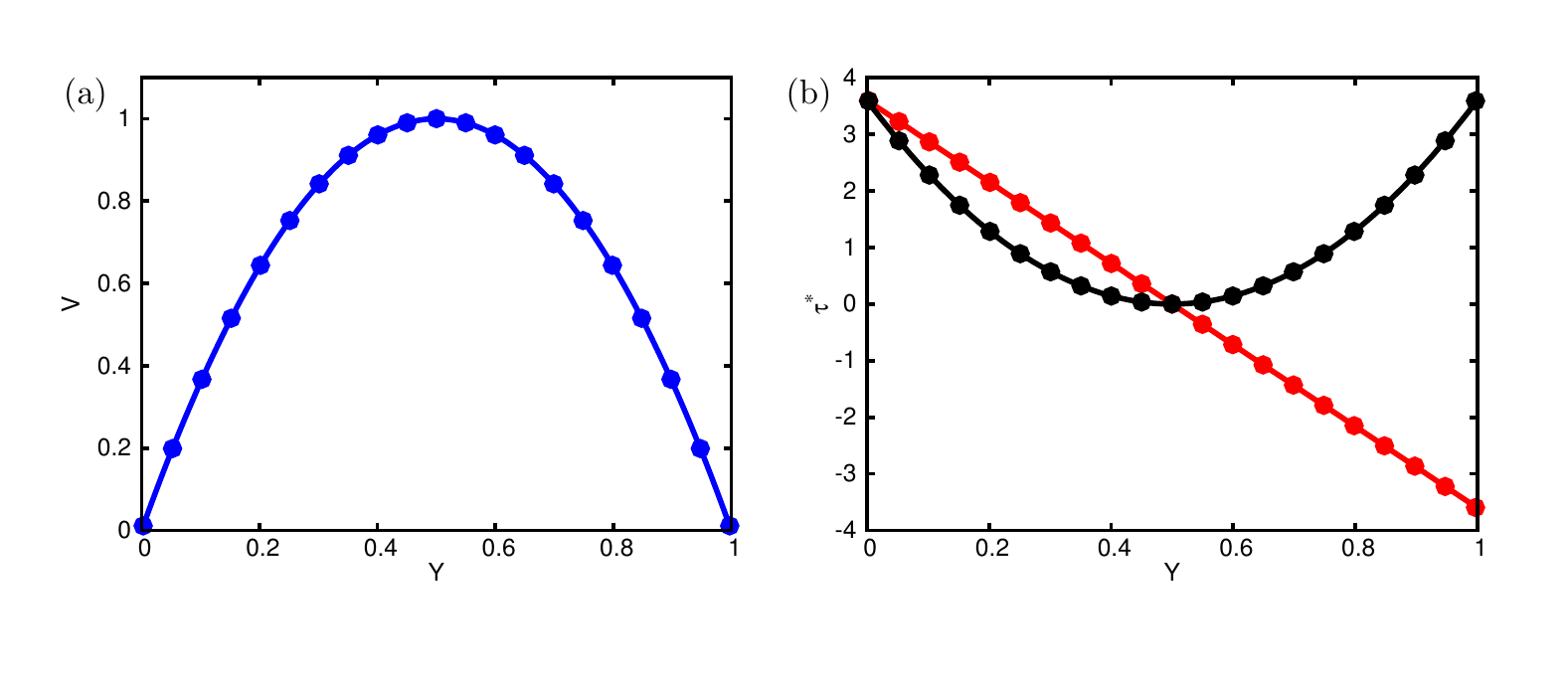}}
\end{center}
\caption{Poiseuille flow of an Oldroyd-B fluid. 
(a) Steady state streamwise velocity component profile, (b) $\tau^*_{xx}$ (black color) and $\tau^*_{xy}$ (red color) stress profiles. 
The symbols represent our numerical results while the solid lines are the analytical solution derived by Waters and King \cite{Waters_RA_1970}. ($Re=0.125$, $Wi=0.125$ and $\beta=0.1$).}
\label{supfigOLD}
\end{figure}

\begin{figure}[htbp]
\begin{center}
{\includegraphics[width=0.9\textwidth]{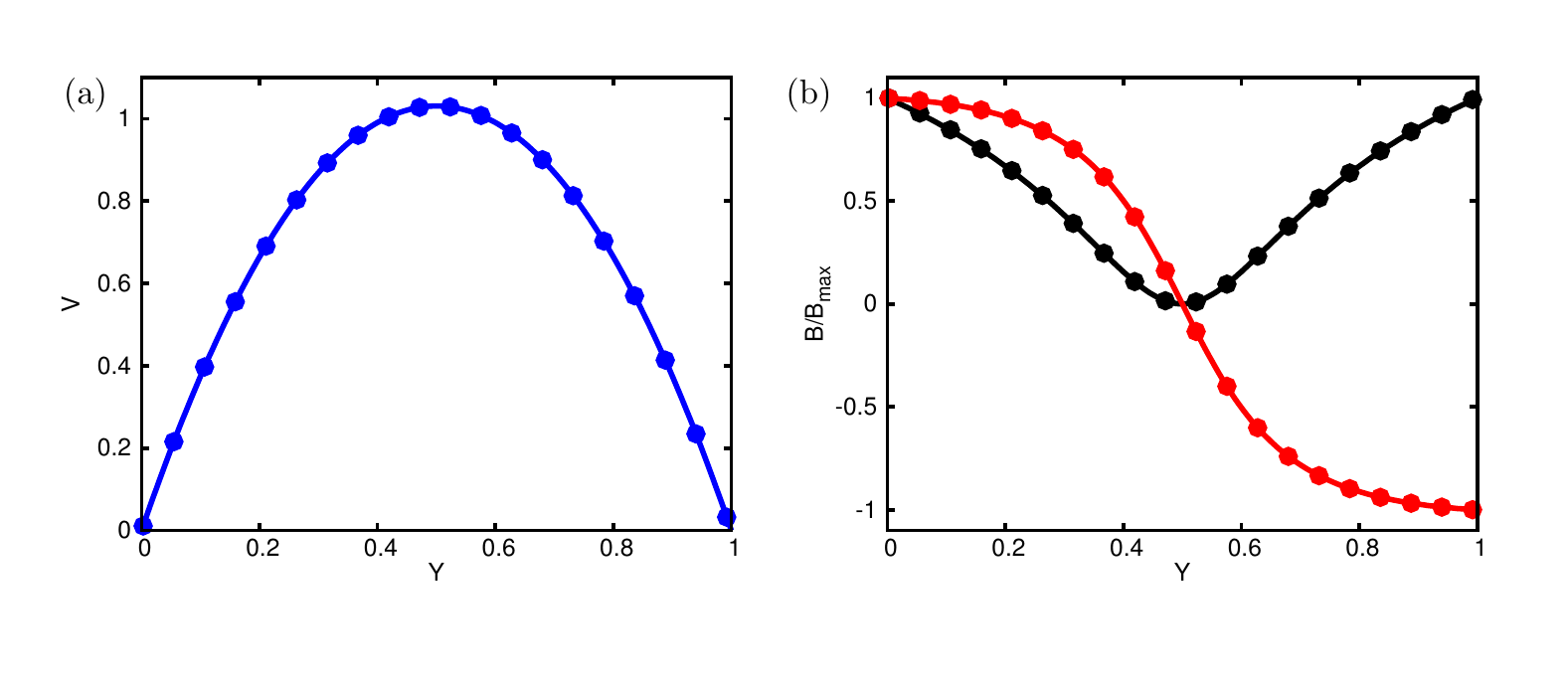}} 
\end{center}
\caption{Poiseuille flow of a FENE-P fluid. (a) Steady state streamwise velocity component profile, (b) $B_{xx}$ (black color) and $B_{xy}$ (red color) profiles. 
The symbols represent our numerical results while the solid lines are the analytical solution. ($Re=300$, $Wi=25$ and $\beta=0.9$).}
\label{supfigFNP}
\end{figure}

\section{Validation }
\label{sec: Validation}
\subsection{Single-phase flow}

\subsubsection{Poiseuille flow of a viscoelastic fluid}
The first test case deals with the start-up Poiseuille flow of an Oldroyd-B and FENE-P fluid ($Bi=0$) in a planar channel. The geometry is a two-dimensional channel bounded by two parallel walls separated by a distance $h=L_y$, where $y$ denotes the wall-normal direction, and $x$ is the streamwise direction. The fluid is initially at rest and set into motion by applying a sudden constant pressure gradient in the 
streamwise direction. No-slip boundary conditions are applied at the walls. As our method solves the three-dimensional Navier-Stokes equations, we impose periodic boundary conditions in the streamwise and 
spanwise directions to emulate the two-dimensional geometry. The following dimensionless variables are introduced: 
 \begin{equation}
Y = \frac{y}{h}; \, \tau^* = \frac{\tau h}{u_0 (\mu_s+\mu_p)}; \, T = \frac{t(\mu_s+\mu_p)}{\rho h^2};\, V=\frac{u}{u_0};
 \, \beta =\frac{\mu_s}{\mu_s+\mu_p};\, Re =\frac{\rho u_0 h}{\mu_s+\mu_p}; \, 
  Wi =\frac{\lambda u_0}{h},
  \label{PRESS}
\end{equation}
where $\tau^*$ is a non-dimensional stress, $T$ is a non-dimensional time and the velocity scale is $u_0=-h^2dp/dx/8(\mu_s+\mu_p)$. The uniform grid has the grid size $\Delta y=h/180$. The time-evolution of the centerline velocity and the wall shear stress is shown for 
the Oldroyd-B fluid case in Fig. \ref{supfig1}. The velocity and the stress components show oscillating behaviour with overshoots and undershoots before settling down to their fully developed values. 
The steady state profiles for velocity and stress components are shown in Figs. \ref{supfigOLD} and \ref{supfigFNP} for the Oldroyd-B and FENE-P fluids, respectively. As can be seen from these figures, there is
an excellent agreement between our numerical and existing analytical results.

\begin{figure}
\centering
\includegraphics[width = 0.9\textwidth]{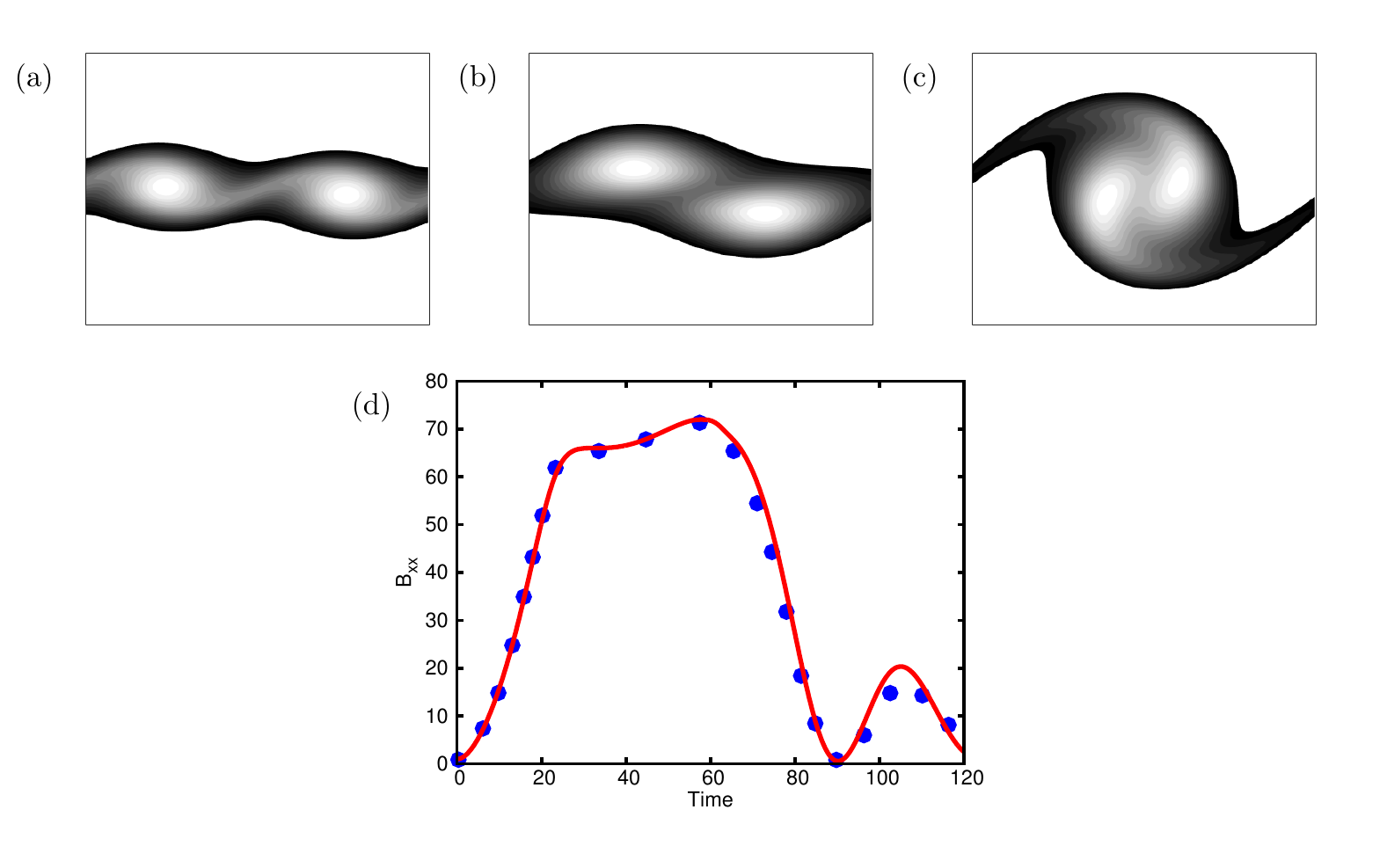}
\caption{(a-c) Instantaneous contours of the absolute value of vorticity at time $T \approx 20$, $60$ and $100$. The color scale from black to white ranges from $0.05$ to $0.4$ in (a), from $0.05$ to $0.3$ in 
(b) and from $0.05 $ to $0.25$ in (c). (d) Time evolution of the  $B_{xx}$ component of the polymer conformation tensor. The red line displays our numerical results, while the symbols display the results of 
Min \etal \cite{Min_JNNFM_2001}. ($Re=50, Wi=25, \beta=0.9, L^2 = 100$).}
\label{fig:val-pol}
\end{figure}

\subsubsection{Temporally evolving mixing layer of a viscoelastic fluid}

The FENE-P model implementation has been validated by simulating a viscoelastic temporally evolving mixing layer flow and by comparing our results with those provided by Min \etal \cite{Min_JNNFM_2001}. 
We consider the initial velocity field $u = 0.5 (\tanh y)$, and trigger the roll-up of the shear layer with a small $2D$ perturbation. 
The characteristic velocity and length scales are 
$\Delta u= u_\textrm{max} - u_\textrm{min}$ and $\delta = \Delta u/(du/dy)_\textrm{max}$, respectively. The Reynolds number is fixed at $Re=\rho \delta \Delta u/\mu_s=50$ and the Weissenberg number at
$Wi=\lambda \Delta u /\delta=25$; moreover, the extensibility $L^2$ is set to $100$, and the solvent viscosity ratio tp $\beta=0.9$. The dimensionless time is defined as $T=t \Delta u/\delta$. 
The $2D$ numerical domain has the size $30 \delta \times 100 \delta$, discretised by $128 \times 384$ grid points. Note that the flow configuration and domain are the same used by Min \etal \cite{Min_JNNFM_2001}. 
Figures \ref{fig:val-pol}(a - c) show the instantaneous vorticity contours for the Newtonian flow, where we can observe that the initial perturbation grows in time and generates two vortices 
(panel a - $T \approx 20$), which subsequently roll up (panel b - $T \approx 60$) and eventually merge into one large vortex (panel c - $T \approx 100$); the polymeric flow shows a similar behavior. 
The quantitative validation is shown in the bottom panel, where we plot the time history of $B_{xx}$ in the center of the domain: the symbols represent the literature results, whereas the red line indicates  our numerical data. We find a good agreement of the conformation tensor component time history over the whole vortex merging process.

\begin{figure}[htbp]
\begin{center}
\begin{tabular}{cc}
{\includegraphics[width=0.9\textwidth]{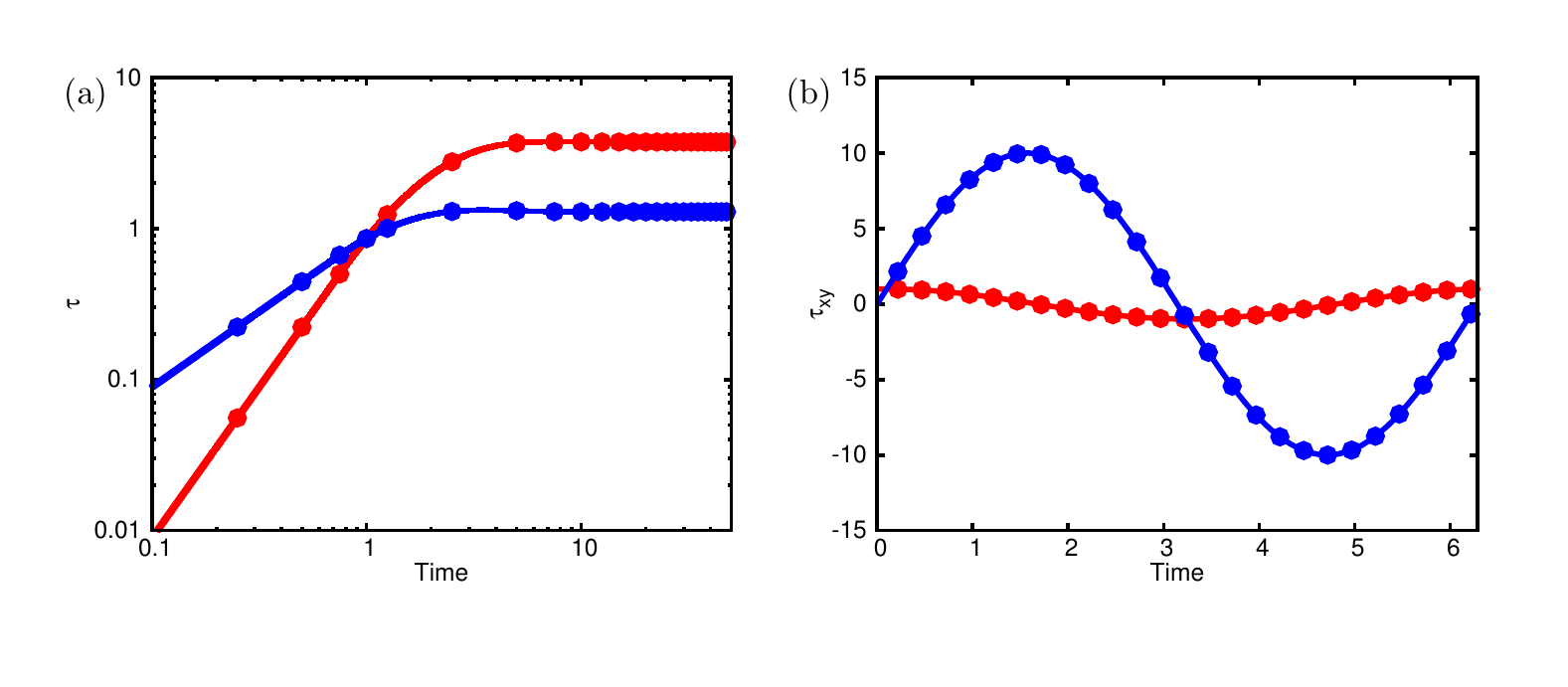}} 
\end{tabular}
\end{center}
\caption{Simple and oscillating shear flow. (a) The evolution of $\tau_{xx}$ (red) and $\tau_{xy}$ (blue) for simple shear flow. (b) The evolution of the shear EVP stress for an oscillating shear flow at $Bi=0$
(red) and $300$ (blue). The solid lines represent the analytical solution by Saramito \cite{saramito2007new} while the symbols are our numerical results. (Simple shear flow: $Bi=1$, $Wi=1$ and $\beta=1/9$. 
Oscillating shear flow: $Wi=0.1$ and $\beta=0$).}
\label{uncoupl}
\end{figure}

\subsubsection{Shear flow of an elastoviscoplastic fluid}
Next, the method is validated for elastoviscoplastic (EVP) single-phase flows. For this purpose, two test cases are considered. The first case is a simple shear flow.
 Initially, the fluid is at rest and set into 
motion by a constant shear rate  $\dot{\gamma}_0$. This test case has a constant dimensionless velocity gradient $\nabla {\bf u}=\bigl[\begin{smallmatrix} 0&1 \\ 0&0 \end{smallmatrix} \bigr]$, 
the Weissenberg number $Wi=\lambda \dot{\gamma}_0=1$, the Bingham number $Bi=\tau_0/\mu_0 \dot{\gamma}_0=1$, and the viscosity ratio $\beta=1/9$. The time evolution of the stresses is shown in Fig. \ref{uncoupl}(a). The stress components increase as long as the yield criterion is not satisfied; once the criterion is fulfilled (T $\approx$ 1), the energy starts to dissipate as a result of viscous effects, which is clearly seen in the figure as the slope of the time evolution of the stresses decreases significantly.

The second test case considers the periodic shear flow of an EVP fluid. An oscillatory flow is applied by imposing the shear strain $\gamma_0 sin(\omega t)$, where $
\gamma_0$ is the strain amplitude and $\omega$ is the angular frequency of the oscillation. 
The Weissenberg number is defined as $Wi=\lambda \omega$ and the Bingham number as $Bi= \tau_y/(\rho \gamma_0 \omega)$ in this case. Computations are performed for two  different Bingham numbers, i.e., $Bi=0$ and $300$. Note that these two values are extreme cases for which the material behaves like a viscoelastic fluid ($Bi=0$) and like an elastic solid ($Bi=300$).  The viscoelastic case can be reached at large strain amplitudes $\gamma_0 \rightarrow \infty$, whereas the elastic solid behavior is obtained when the amplitude is small $\gamma_0 \rightarrow 0$. 
The other dimensionless parameters of the problem are kept constant at $Wi=0.1$ and $\beta=0$. The evolution of the shear stress component $\tau_{xy}$ is displayed in Fig. \ref{uncoupl}(b) for $Bi=0$ and $300$. As can be seen in these figures, there is a good agreement between our simulation and the analytical solutions, thus indicating an accurate solution of the EVP model equations.

\subsection{Multiphase flow in complex fluids}

\begin{table}[b]
\caption{Comparison of the settling velocity of a sphere in EVP fluid.}
\centering
\begin{tabular}{ c c c }
  \hline			
   & $V$ (mm s$^{-1}$) \\
  \hline			
  Present work & 0.356  \\
  Fraggedakis \etal \cite{Fraggedakis_SM_2016} & 0.364 \\
  Holenberg \etal \cite{holenberg2012particle} & 0.37 \\
  \hline  
\end{tabular}
\label{Frag}
\end{table}

\begin{figure}[th]
\begin{center}
{\includegraphics[width=0.9\textwidth]{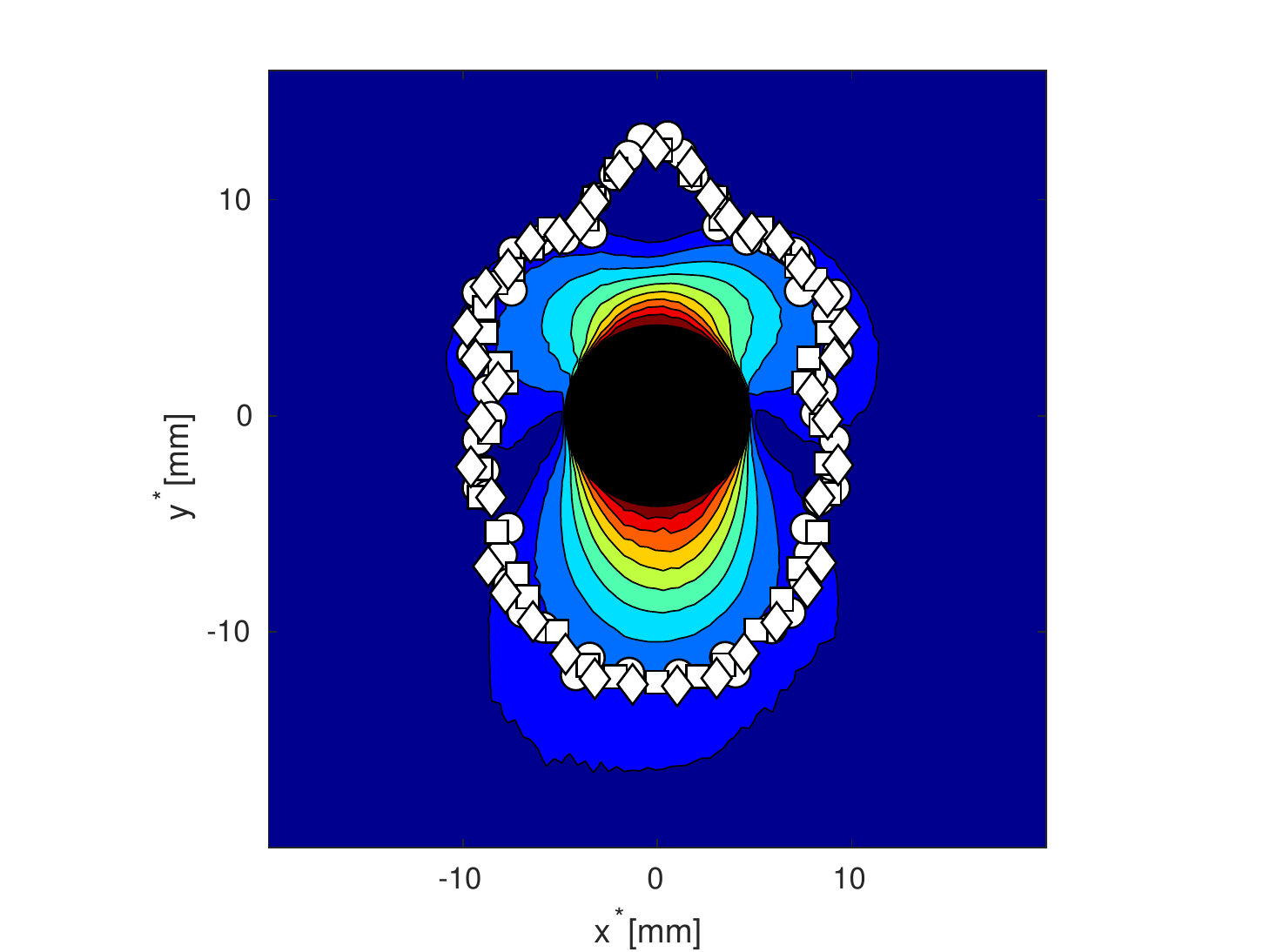}}
\end{center}
\caption{Sphere settling in EVP fluid. Velocity magnitude, scaled with the terminal velocity of the settling sphere. The different symbols represent different experimental series for the solid-fluid boundary defined by Holenberg \etal \cite{holenberg2012particle}. 
($Ar=0.03$, $Wi=1.04$, $Bn=0.089$, $\rho^\circ=1.38$ and $\beta=0.01$).}
\label{str1}
\end{figure}

\subsubsection{Sedimentation of a spherical particle in an elastoviscoplastic fluid}
After validating the numerical method for simple viscometric flows, the method is now applied to study the sedimentation of a spherical particle in a channel filled with an EVP 
fluid. This problem exhibits different viscometric flows simultaneously, i.e., biaxial stretching upstream of the particle, shear flow on the sides, and uniaxial extensional flow downstream of it. 
The Saramito model is employed here to facilitate comparison of the present results 
with the numerical results by Fraggedakis \etal \cite{Fraggedakis_SM_2016} and with the experimental data by Holenberg \etal \cite{holenberg2012particle}. 
A single spherical particle of radius $R$ is centered in a domain of size $(L_x \times L_y \times L_z)=(12R \times 20R \times 12R)$; a grid of $144\times240\times144$ points is used to discretize the computational 
domain. Periodic boundary conditions are imposed in the $x$ (spanwise) and $y$ (gravity) directions whereas a free slip/no penetration condition is enforced in the $z$ direction.  
The particle starts moving due to the gravity in an otherwise quiescent ambient EVP fluid. Following Fraggedakis \etal \cite{Fraggedakis_SM_2016}, the non-dimensional parameters are defined as follows: 
 \begin{equation}
Ar = \frac{\Delta \rho^2 g R^3}{\mu_s+\mu_p}; \quad Wi=\frac{\lambda \Delta \rho g R}{\mu_s+\mu_p};
\quad Bn = \frac{\tau_y}{\Delta \rho g R}; \quad \rho^\circ=\frac{\rho_s}{\rho_f}, \label{NDFrag}
\end{equation}
where $Ar$, $Wi$, $Bn$, and $\rho^\circ$ represent the Archimedes number, the Weissenberg number, the Bingham number, and the density ratio, respectively. In Eq.\ (\ref{NDFrag}), the density difference is defined 
as $\Delta \rho = \rho_f (\rho^\circ-1) = \rho_s-\rho_f$, where $\rho_s$ and $\rho_f$ are the solid and the fluid densities, respectively. 
The present results are compared with the computational simulations by Fraggedakis \etal \cite{Fraggedakis_SM_2016} and the experimental results by Holenberg \etal \cite{holenberg2012particle}. 
The simulation is performed for $Ar=0.03$, $Wi=1.04$, $Bn=0.089$, $\rho^\circ=1.38$ and $\beta=0.01$. Note that the ``rough heavy sphere" case is considered in the present study, referring 
to the no-slip boundary condition on the particle. First, a quantitative comparison is conducted based on the steady state settling speed of the particle in the EVP fluid. The terminal velocity 
predicted by the present study, the one in the numerical simulation in Ref. \cite{Fraggedakis_SM_2016} and the one observed experimentally \cite{holenberg2012particle} are reported in Table \ref{Frag}. 
As can be seen from the table, the present results could capture the terminal velocity accurately, and indeed the result of the present study deviates from that in other published studies by  less than $4\%$. 
Next, we present a qualitative comparison of the velocity fields around the spherical particle. The steady state velocity contours normalized with the particle terminal velocity are displayed in 
Fig. \ref{str1}. 
Direct comparisons with the experimental data of \cite{holenberg2012particle} 
are also included for the yield surface represented by the white markers around the sphere, where the circular and square marks indicate two different experimental series. 
Holenberg \etal \cite{holenberg2012particle} 
determined the yielded region by means of PIV and PTV techniques, i.e., they defined the yielded region where the velocity magnitude exceeded $10\%$ of the settling velocity. 
To facilitate direct comparison with experimental data, the velocity contours shown in Fig. \ref{str1} are constructed as follows: the distance between the consecutive contour lines is the same and 
equals to $10\%$ of the terminal velocity, starting from $10\%$ to $90\%$ of the velocity. 
Generally, we are in good agreement with the experimental marks \cite{holenberg2012particle} and 
simulation results shown in Fig. 9 in the work of Fraggedakis et al. \cite{Fraggedakis_SM_2016}. Similarly to \cite{Fraggedakis_SM_2016}, the current methodology could capture the expected loss of the fore-aft symmetry. On the other hand, a slight discrepancy 
between the present results and those by the aforementioned work \cite{Fraggedakis_SM_2016} can be attributed to a different computational box and a lower resolution. Indeed, in the present study a full three-dimensional 
flow is employed, whereas Fraggedakis \etal \cite{Fraggedakis_SM_2016} consider an axisymmetric configuration. Moreover, local grid refinement is used in Ref.~\cite{Fraggedakis_SM_2016}. 
Their very fine grid in the vicinity of the sphere may result in an improved resolution of the yielded region. 
Another reason could be the employment of different boundary conditions in the far-field boundary, the open-boundary condition is used by Fraggedakis \etal \cite{Fraggedakis_SM_2016} 
while a periodic boundary condition is employed in the present work.

\begin{figure}[t]
\begin{center}
{\includegraphics[width=0.9\textwidth]{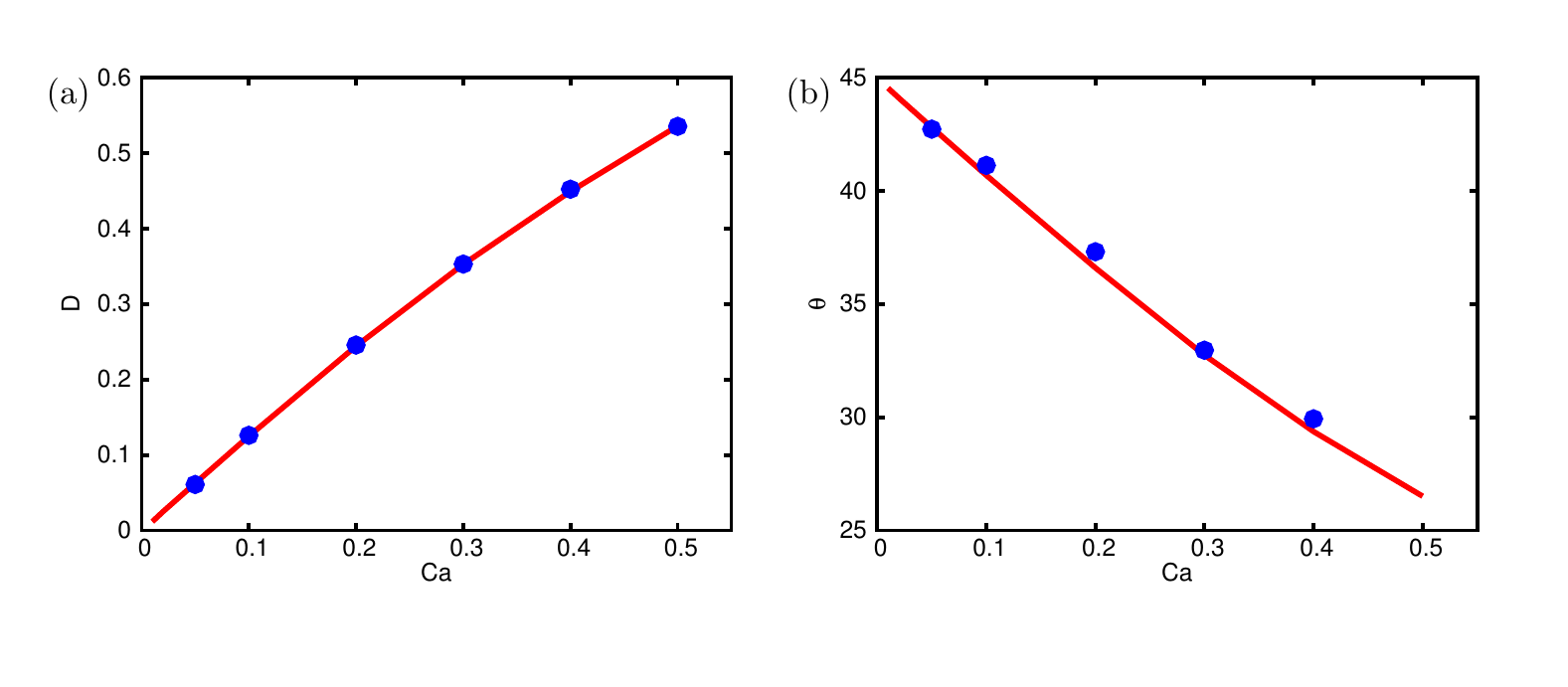}} 
\end{center}
\caption{Steady deformation of a neo-Hookean elastic particle in a Newtonian fluid for $0.05 \le Ca \le 0.5$. (a) Taylor parameter $D$ and (b) inclination angle $\theta$ vs. $Ca$. Red line: numerical results from Ref.~\cite{Villone_CF_2014};  blue circles: our numerical simulation. ($Re=0.1$).}
\label{edfig1}
\end{figure}

\subsubsection{Deformable dilute suspension in a shear flow} 

\paragraph*{Steady deformation of a neo-Hookean elastic particle in a shear flow} \mbox{}\\
In this test case, we simulate the flow in a plane Couette geometry. We use a Cartesian uniform mesh in a rectangular box of size $16R \times 10R \times 16R$, 
with 16 grid points per particle radius $R$. Periodic boundary conditions are imposed in the streamwise ($x$) and spanwise ($z$) directions, and the no-slip condition at the walls  ($y = -h$ and $y = h$), which move in two opposite directions with a constant streamwise velocity $\pm U = h \dot{\gamma}$. The Reynolds number $Re=\rho\dot{\gamma}R^2/\mu$ is fixed at $0.1$ and the Capillary number $Ca=\mu \dot{\gamma}/G$ varied one order of magnitude between $0.05$ and $0.5$. After the transients die out, the sphere deforms to approximately an ellipsoid, and we therefore characterize these shapes 
by the Taylor parameter ($D$) and the angle $\theta$. The Taylor deformation parameter is defined as $D=(L-B)/(L+B)$, where $L$ and $B$ are the major and 
minor axis of the equivalent ellipsoid in the middle plane, and $\theta$ is the inclination angle with the respect to the streamwise direction.
The steady state values of $D$ and $\theta$ are reported in Fig. \ref{edfig1} for different $Ca$, and compared with those by 
Villone \etal \cite{Villone_CF_2014}. Similarly to the case of a viscoelastic droplet in a Newtonian medium, deformation as well as the tendency to align with the flow  increases with $Ca$, 
i.e. with the deformability.  A very good agreement is found between our numerical results and those in the literature. Further validation and details of our implementation can be found in 
\cite{Rosti_JFM_2017,rosti1,rosti2}.

\begin{figure}[t]
\begin{center}
{\includegraphics[width=0.9\textwidth]{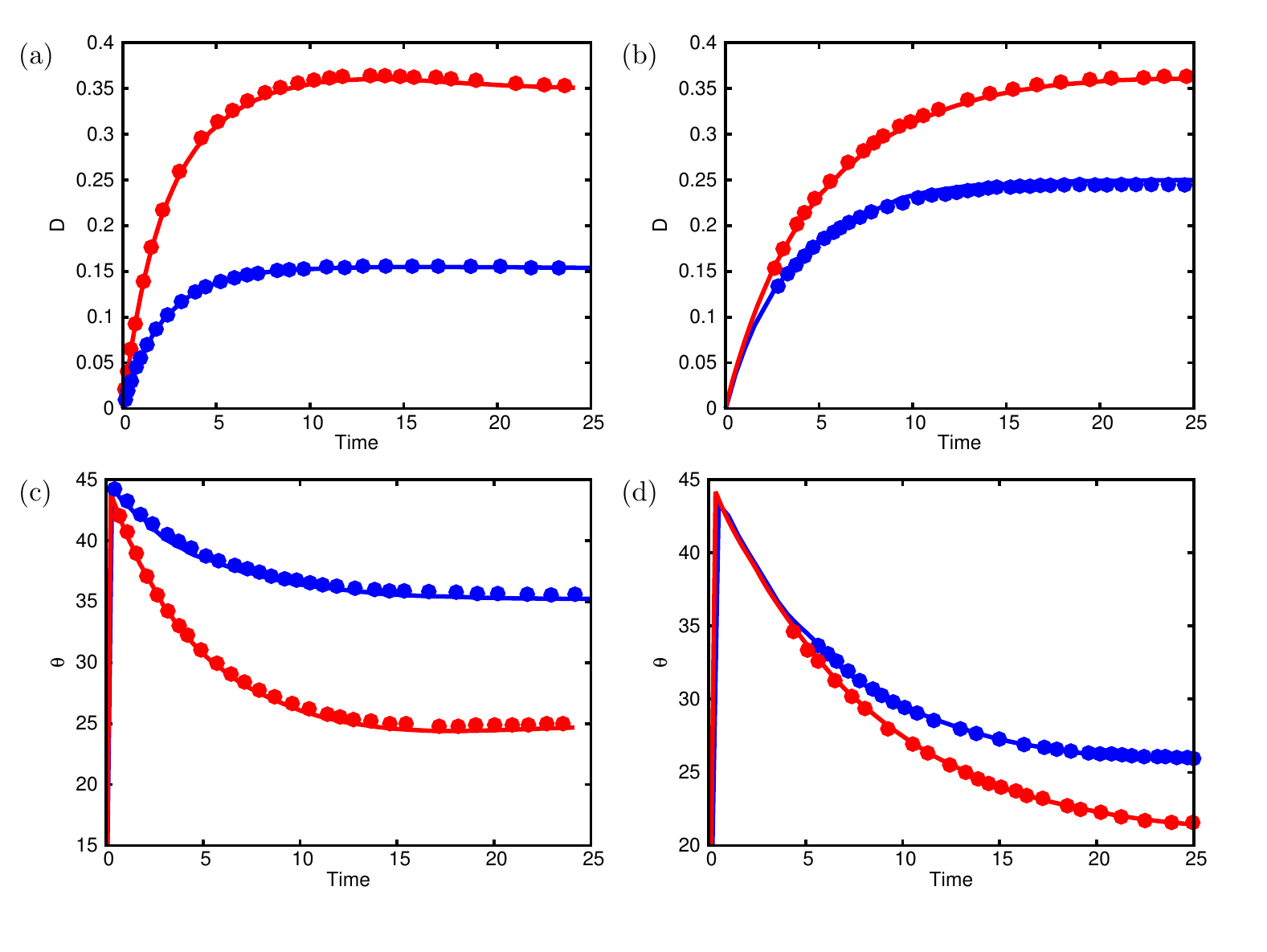}} 
\end{center}
\caption{Drop deformation under simple shear flow for viscoelastic droplet in Newtonian medium (VN) and for a Newtonian droplet in a viscoelastic medium (NV). 
The solid lines represent our results while the symbols are those in Ref.~\cite{Verhulst_JNNFM_2009} and \cite{Cardinaels_JNNFM_2011} for the VN (a and c) and NV (b and d) systems, respectively. 
Panels (a) and (b) show the temporal evolution of the deformation $D$, and the (c) and (d) the history of the angle $\theta$. 
For the VN case, $Ca=0.14$ (blue) and $0.32$ (red) and for the NV case, $\chi=0.46$ (blue) 
and $\chi=0.76$ (red).  (VN case: $Re=0.05, De_2=1.54, \beta=0.68, k_{\rho}=1, k_{\mu}=1.5$ and $\chi=0.25$; NV case: $Re=0.1, De_1=1, Ca=0.2, \beta=0.68, k_{\rho}=1$ and $k_{\mu}=1.5$).}
\label{sdfig1}
\end{figure}

\paragraph*{Three-dimensional viscoelastic droplet} \mbox{}\\
Verhulst \etal \cite{Verhulst_JNNFM_2009} and Cardinaels \etal \cite{Cardinaels_JNNFM_2011} considered fully three-dimensional shear-driven droplets in which either the droplet or the surrounding fluid is viscoelastic. The Oldroyd-B model is employed in the present study to facilitate a direct comparison with the results of Verhulst \etal \cite{Verhulst_JNNFM_2009} and Cardinaels \etal \cite{Cardinaels_JNNFM_2011}. 
The spherical droplet of radius $R$ is at the center of the computational domain. Opposite velocities,  $V$ and $-V$, are enforced on the two walls located at $z = 0$ and $z = H$
 to obtain the shear rate $\dot{\gamma}=2V/H$. Periodic boundary conditions are imposed in the $x$ (spanwise) and $y$ (streamwise) directions and no-slip conditions at the two walls. 
Following Verhulst \etal \cite{Verhulst_JNNFM_2009} and Cardinaels \etal \cite{Cardinaels_JNNFM_2011}, the non-dimensional parameters are defined as follows: the Reynolds number $Re=\rho_1 \dot{\gamma}R^2/\mu_1$, 
the capillary number $Ca=R \dot{\gamma} \mu_1/\sigma$, the Weissenberg number $Wi=\lambda \dot{\gamma}$, the viscosity ratio $k_{\mu}=\mu_2/\mu_1$, the density ratio $k_{\rho}=\rho_2/\rho_1$, and the confinement ratio 
$\chi=2R/H$. Alternatively, two Deborah numbers can be defined as $De_1=(1-\beta)Wi/Ca$ and $De_2=(1-\beta)Wi/(k_{\mu} Ca)$. 
The results are presented in terms of the Deborah number and dimensionless 
capilary time $t \dot{\gamma}/Ca$. The droplet deformation in the $y-z$ plane is measured by the Taylor deformation parameter introduced above.
Following Ramanujan and Pozrikidis \cite{Ramanujan_JFM_1998}, the inertia tensor of the drop is used to find the equivalent 
ellipsoid. 

First, we consider the startup dynamics of an Oldroyd-B droplet in a Newtonian medium (VN). 
The viscoelastic spherical droplet is centred in a computational domain of size $L_x=H, L_y=2H, L_z=H$, 
which is discretised with a resolution of $\Delta x= \Delta y = \Delta z = H/192$. The simulations are performed at $Re=0.05, De_2=1.54, \beta=0.68, k_{\rho}=1, k_{\mu}=1.5$ and $\chi=0.25$. 
The time evolutions of the Taylor parameter and the angle of inclination for a viscoelastic droplet in a Newtonian fluid at $Ca=0.14$ and $0.32$ are depicted in Fig. \ref{sdfig1}(a) together with the 
numerical results by Verhulst \etal \cite{Verhulst_JNNFM_2009}. As expected, the drop deformation and alignment with the flow increase with $Ca$. Also, the time evolution of both the Taylor parameter 
and the inclination angle are in good agreement with the results reported in Ref.~\cite{Verhulst_JNNFM_2009}. 

Next, the dynamics of a Newtonian droplet in an Oldroyd-B fluid (NV) is studied. The resolution is fixed at $\Delta x= \Delta y = \Delta z = H/64$ and the computational domain is $L_x=2H, L_y=4H, L_z=H$. 
The computations were performed for $Re=0.1, De_1=1, Ca=0.2, \beta=0.68, k_{\rho}=1$ and $k_{\mu}=1.5$. The time evolution of the drop deformation parameter and its orientation angle are shown in 
Fig. \ref{sdfig1}(b) for two different confinement ratios: $\chi=0.46$ and $\chi=0.76$. As can be seen in the figure, the confinement ratio increases both the drop deformation and the drop orientation angle. 
The comparison between the present results and those by Cardinaels \etal \cite{Cardinaels_JNNFM_2011} shows good agreement.

\subsubsection{Buoyancy-driven droplet in viscoelastic and elastoviscoplastic media}
Finally, the method is validated for buoyancy-driven (rising) droplets. We start from a Newtonian droplet rising in a Newtonian and viscoelastic fluid.
The Oldroyd-B model is used in the present work to facilitate direct comparison with the results by Prieto \cite{Prieto_JNNFM_2015}, Zainali \etal \cite{Zainali_CMAME_2013} 
and Vahabi and Sadeghy 
\cite{Vahabi_NRG_2015}. The fully Newtonian case is a classical benchmark, see e.g., Hysing \etal \cite{Hysing_IJNMF_2009}. 
The domain is rectangular with the width $L_x=1$ and the height $L_y=2$. A spherical droplet with a radius $R$ is initially placed at the centerline of the channel at a distance of $L_y/4$ from the lower part of the 
channel. The no-slip boundary conditions are applied at the horizontal walls. It should be noted that Prieto \cite{Prieto_JNNFM_2015} used the free-slip boundary conditions on the vertical walls, 
whereas Zainali \etal \cite{Zainali_CMAME_2013} and Vahabi and Sadeghy \cite{Vahabi_NRG_2015} imposed no-slip boundary conditions. The non-dimensional parameters pertaining to this problem are defined as: 
the Reynolds number $Re=\rho_1U_g L/\mu_1$, the E${\rm\ddot{o}}$tv${\rm\ddot{o}}$s number $Eo=\rho_1 U_g^2 L/\sigma$, the Weissenberg number $Wi=\lambda U_g/L$, Bingham number $Bi=\tau_y/\rho g L$, 
the viscosity ratio $k_{\mu}=\mu_2/\mu_1$, the density ratio $k_{\rho}=\rho_2/\rho_1$ and the confinement ratio $\chi=2R/L_x$. The reference length scale is $L=2R$, the velocity scale is $U_g=\sqrt{g L}$, 
where $g$ is the gravitational constant, and the time scale is $L/U_g$. 

\begin{figure}[t]
\begin{center}
{\includegraphics[width=0.9\textwidth]{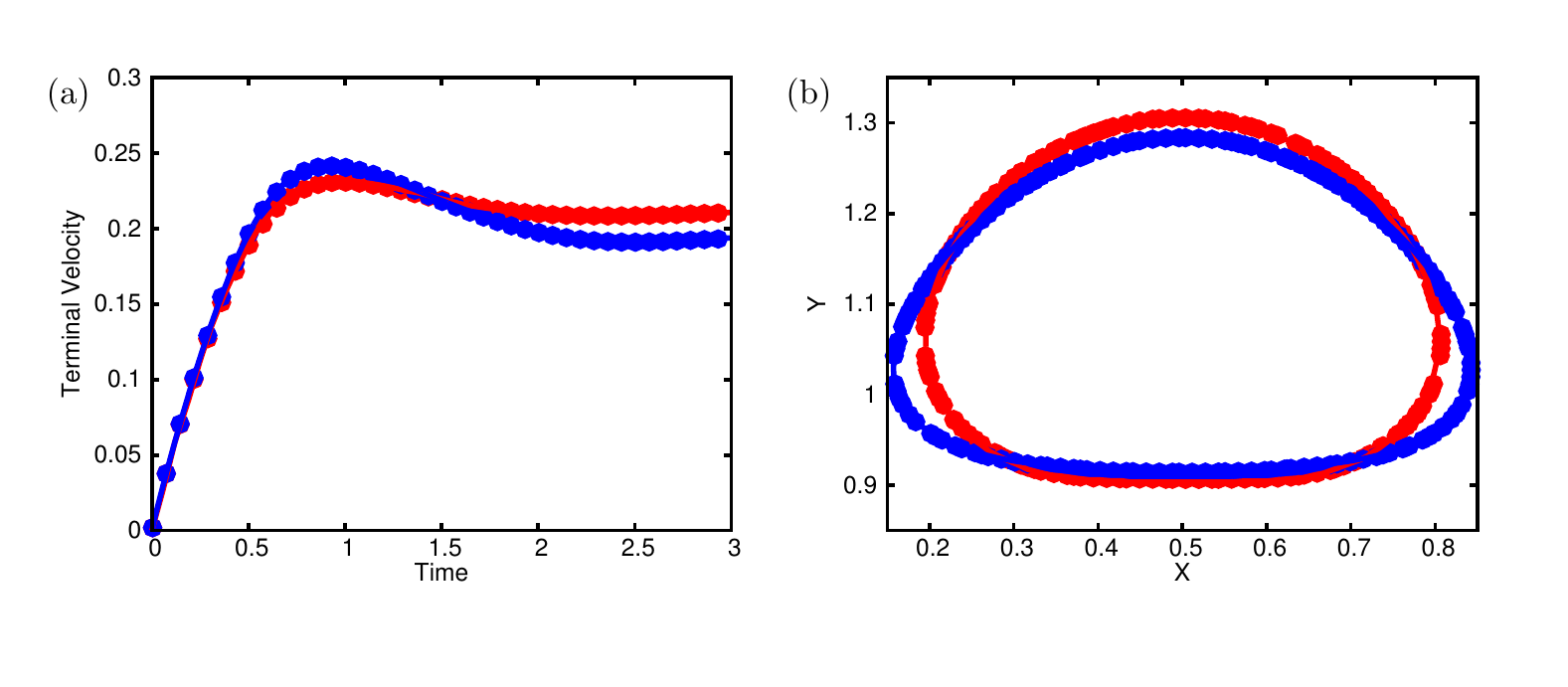}} 
\end{center}
\caption{Buoyancy-driven viscoelastic two-phase system: Newtonian case (blue color) and Newtonian droplet in viscoelastic medium (red color). (a) The terminal velocity versus the non-dimensional time and 
(b) its respective steady state shape. The symbols represent the present results while the solid lines those by Prieto \cite{Prieto_JNNFM_2015}. Our results are obtained using a $128 \times 256$ grid. 
(N case: $Re=35,Eo=10,Bi=0,k_{\rho}=0.1, k_{\mu}=0.1,$ and $\chi=0.5$; NV case: $Re=35, Eo=10, Wi=1,Bi=0, \beta=0.5, k_{\rho}=0.1, k_{\mu}=0.1$ and $\chi=0.5$). }
\label{bdfig1}
\end{figure}
First, we show a comparison of rising droplets to the computational study of Prieto \cite{Prieto_JNNFM_2015} in two cases: (NV) denotes a Newtonian droplet in a viscoelastic medium ($Wi=1$, $\beta=0.5$), 
and (N) denotes a Newtonian droplet in a Newtonian medium ($Wi=0$, $\beta=1$). The other parameters are: $Re=35, Eo=10, Bi=0, k_{\rho}=0.1, k_{\mu}=0.1$ and $\chi=0.5$. Figure \ref{bdfig1} 
shows the evolution of the terminal velocity and the steady state shape for a fully Newtonian (N) case and for a Newtonian drop in a viscoelastic medium (NV), both of which are in good agreement with the 
literature results. Note that, in the study of Prieto \cite{Prieto_JNNFM_2015}, the microscopic Hooke model was used rather than the Oldroyd-B model considered in the present work; despite of that, very similar results are obtained.

\begin{figure}[t]
\begin{center}
\begin{tabular}{cc}
{\includegraphics[width=0.5\textwidth]{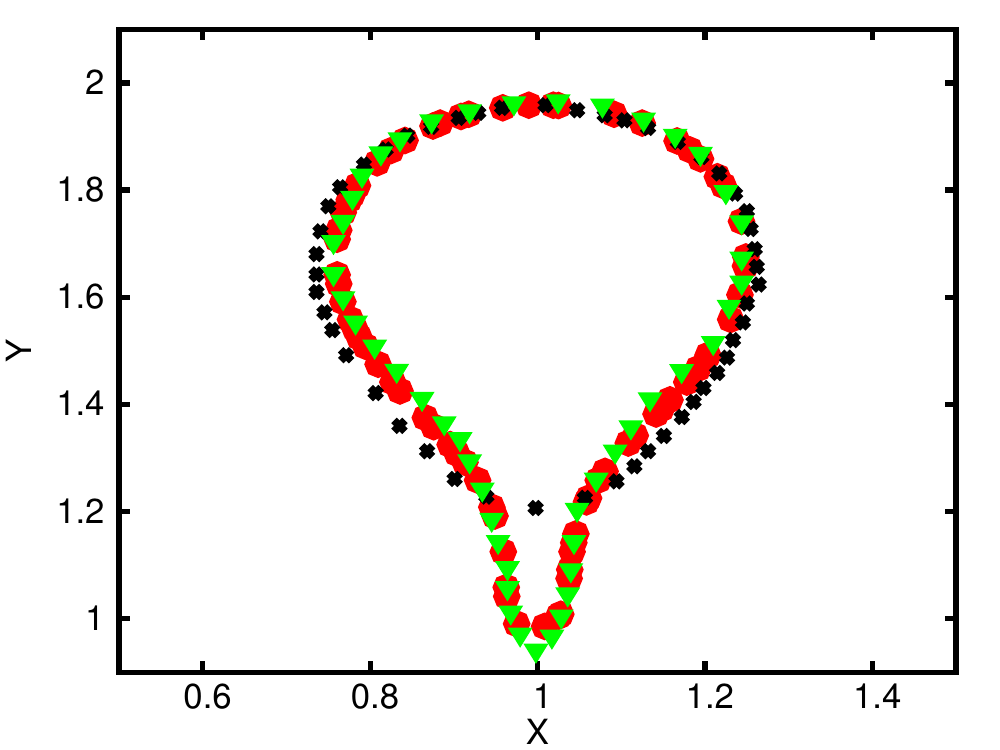}} 
\end{tabular}
\end{center}
\caption{Shape of a Newtonian droplet rising in an Oldroyd-B fluid at $t=0.13s$. The present results ($\circ$) are compared with the results of Zainali \etal \cite{Zainali_CMAME_2013} ($\times$) 
and the results of Vahabi and Sadeghy \cite{Vahabi_NRG_2015} ($\triangledown$). ($Re=1.419, Eo=35.28, Wi=8.083,Bi=0, \beta=0.07, k_{\rho}=0.1, k_{\mu}=0.1, \chi=0.3,$ with $120 \times 240$ grid points).}
\label{bdfig2}
\end{figure}

Next, we compare our results in the NV case against the results by Zainali \etal \cite{Zainali_CMAME_2013} and Vahabi and Sadeghy \cite{Vahabi_NRG_2015}. Following these authors, the values of the non-dimensional parameters are $Re=1.419, Eo=35.28, Wi=8.083, Bi=0, \beta=0.07, k_{\rho}=0.1, k_{\mu}=0.1$ and $\chi=0.3$. The droplet interface shapes that we obtained at $t=0.13s$ together with the ones by Zainali \etal \cite{Zainali_CMAME_2013} and Vahabi and Sadeghy \cite{Vahabi_NRG_2015} are depicted in Fig. \ref{bdfig2}. It can be seen 
in the figure that the present result is consistent with the one reported by Vahabi and Sadeghy \cite{Vahabi_NRG_2015}; on the contrary, Zainali \etal \cite{Zainali_CMAME_2013} have not observed the cusped 
trailing edge, which is however a common feature for the case of Newtonian droplet in viscoelastic medium at high polymer concentrations 
\cite{atarita1965aiche,Pillapakkam_JFM_2007,Izbassarov_JNNFM_2015, Prieto_JNNFM_2015,fraggedakis2016jfm}. 

\begin{figure}[htbp]
\begin{center}
\begin{tabular}{cccc}
{\includegraphics[width=0.9\textwidth]{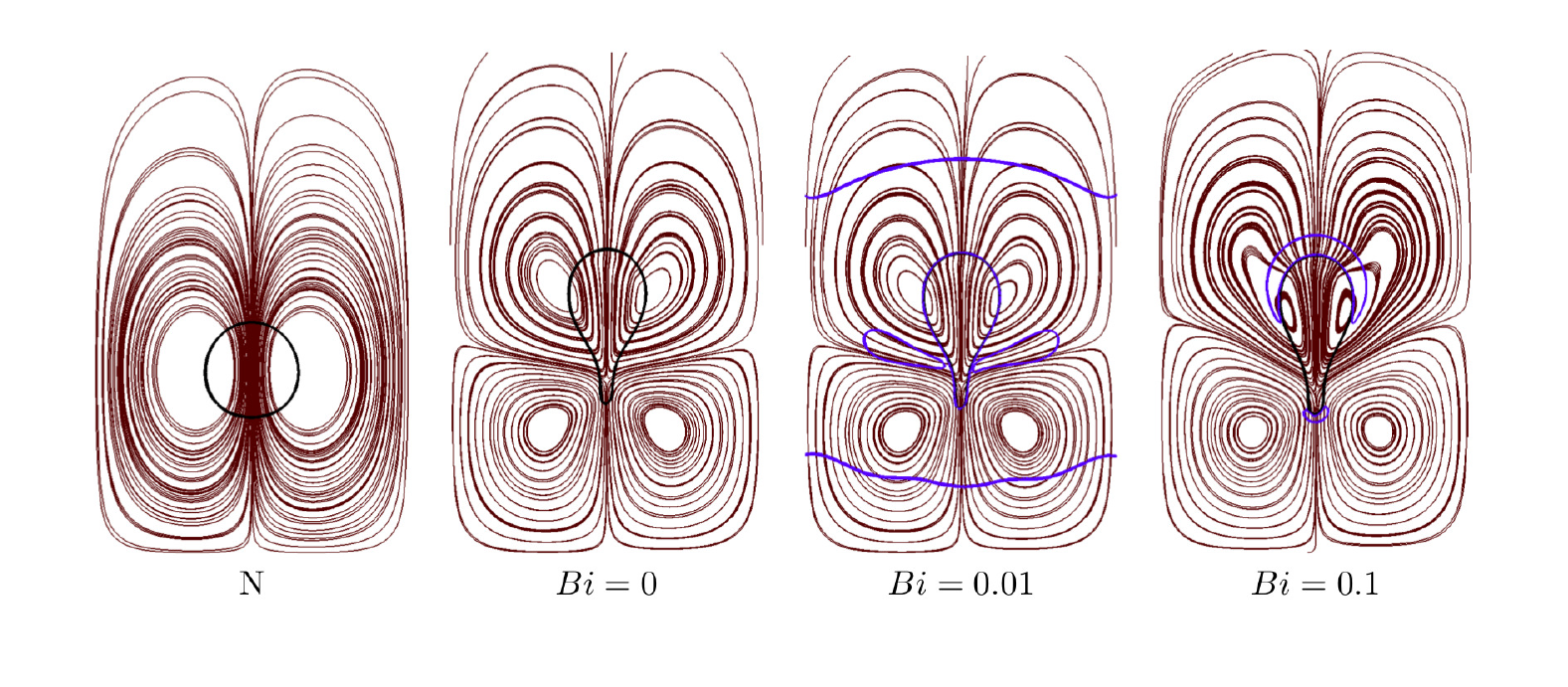}}
\end{tabular}
\end{center}
\caption{Streamlines for the Newtonian droplet rising in a EVP fluid and its respective shape at $t=0.13 s$. Computations are performed for the Newtonian case (N), and EVP fluids with $Bi=0, 0.01$ and $0.1$. 
The blue line denotes the solid-fluid boundary defined via contour $\mathcal{F}=0.5$. ($Re=1.419, Eo=35.28, Wi=8.083, \beta=0.07, k_{\rho}=0.1, k_{\mu}=0.1, \chi=0.3,$ and Grid: $120 \times 240$). }
\label{bdfig3}
\end{figure}

Finally, some sample simulations are presented for a Newtonian droplet moving in an EVP fluid. The physical properties pertinent to the problem are the same as in Zainali \etal \cite{Zainali_CMAME_2013} 
and Vahabi and Sadeghy \cite{Vahabi_NRG_2015} except for a non-zero Bingham number; indeed, the Bingham number is varied between $Bi=0$ and $0.1$. Fig. \ref{bdfig3} shows shapshots at $t=0.13s$ of the 
streamlines inside the computational domain for the fully Newtonian case (N), and for the EVP fluid with $Bi=0, 0.01$ and $0.1$. In the figure, blue line denotes the solid-fluid boundary defined via the isoline $\mathcal{F}=0.5$, where $\mathcal{F}$ is defined in Table \ref{param}.
Note that all the non-Newtonian cases display a negative wake, and therefore they have four closed streamline zones instead of two zones for the Newtonian droplet.
When increasing the Bingham number $Bi$, the extent of the yielded region decreases and the solid-fluid boundary approaches the droplet. 
At $Bi=0.01$, the fluid region occupies most of the domain, but there is a solid region above and below the droplet, as well as two small ellipsoidal regions on both sides of the droplet. 
Finally for $Bi=0.1$, the solid region occupies almost the whole domain, except for two narrow "caps" at the trailing and leading edges of the droplet. 

\section{Conclusion}
An efficient solver has been presented for the three-dimensional direct numerical simulations of viscoelastic and elastoviscoplastic multiphase flows, expected to allow large-scale simulations also in inertial and turbulent regimes. The solver is general and applicable to non-Newtonian fluids with a dispersed phase which is either rigid or deformable (drops, bubbles and elastic particles). The fluid phases can be chosen to be simple EVP fluids following the model of Saramito\cite{saramito2007new}. The method can be later adapted to more complex EVP models. 

To obtain a stable and accurate solution of the transport equations for the stresses (EVP, elastic or viscoelastic), we use  a fifth-order upwinded WENO scheme for the advection term in the stress model equations. This is found to be very robust and considerably less expensive than the third-order compact upwind scheme suggested in the literature. To avoid numerical breakdown at moderate Weissenberg numbers, a local artificial diffusion can be added. We find that a local diffusion is preferable to the global diffusion which can lead to inaccurate solutions by significantly smearing out the gradients.

The interface between the continuous and dispersed phases is tracked using different approaches for different systems. For the case of deformable viscoelastic particles, we adopt an indicator function based on the local volume fraction. For droplets, we utilise a mass-conserving level set method recently developed by this group, including an accurate computation of the surface tension force based on the local curvature, and a highly efficient and scalable FFT-based pressure solver for density-contrasted flows. The overall solution approach proposed here is independent of the specific interface tracking method. The advantage of these methods is that they are fully Eulerian, efficient, accurate and portable from existing available implementations. For rigid particles, on the other hand, the interface is tracked using an immersed boundary method. In this case, the carrier phase is solved on a fixed Eulerian grid, whereas the interface is represented by a Lagrangian grid following the particle. When comparing to the conventional body fitted grid, the IBM is more simple and versatile for moving rigid bodies.

The method is first validated for single-phase elastoviscoplastic flows including the start up flow in planar channel, temporally evolving mixing layer and simple and oscillating shear flows. Then, it is  applied to the sedimentation of a spherical particle in an EVP fluid, a viscoelastic drop under shear flow and a buoyancy-driven viscoelastic droplet. In all the cases mentioned above, the results obtained with our code are found to be in good agreement with previous results found in the literature. Finally, sample results are presented for a Newtonian droplet rising in an elastoviscoplastic fluid. This, and the behaviour of rigid particle suspensions in EVP fluids, will be interesting topics for future investigations.

The present methodology can also handle multi-body issues. For solid particles, we have a soft-sphere collision model \cite{Costa2015} and lubrication corrections \cite{Ardekani2016}
for short-range particle-particle and particle-wall interactions. In particular, when the gap width between two particles (or particles and wall) reduces to zero,
a soft sphere collision model is activated, to calculate the normal and tangential collision force. We will extend the work on collision models to non-Newtonian fluids in the future. In the level-set method, coalescence takes place automatically. However in some cases, this phenomenon needs to be prevented.
In our previous work \cite{Ge_AX_2017}, a hydrodynamic model was derived for the interaction forces induced by depletion of surfactant micelles. As a future study,
this model could be extended to take into account other effects of surfactants, such as diffusion at the interface and in the bulk fluid.
\label{sec: conclusion}

\section*{Acknowledgments}
We acknowledge financial support by the Swedish Research Council through grants No. VR2013-5789, No. VR 2014-5001 and VR2017-4809. This work was supported by the European Research Council Grant no. ERC-2013-CoG-616186, TRITOS and by the Microflusa project. This effort receives funding from the European Union Horizon 2020 research and 
innovation programme under Grant Agreement No. 664823.  S.H. acknowledges financial 
support by NSF  (Grant No. CBET-1554044-CAREER), NSF-ERC (Grant No. CBET-1641152 Supplementary CAREER) and ACS PRF (Grant No. 55661-DNI9). The authors acknowledge computer time provided by SNIC 
(Swedish National Infrastructure for Computing) and OSC (Ohio Supercomputer Center).

\bibliography{IJNMF-FLD-18-0051R1}%
\end{document}